\documentclass[]{jfm}

\usepackage[utf8]{inputenc}
\usepackage{graphicx}
\usepackage{amsmath,amssymb}
\usepackage{multicol}
\usepackage{natbib}
\usepackage{color}
\usepackage{lineno}
\usepackage{bm}

\begin{document}

\newtheorem{lemma}{Lemma}
\newtheorem{corollary}{Corollary}
\renewcommand{\vec}[1]{\mbox{\boldmath$ #1 $}}

\shorttitle{Rheology of a monolayer of squirmers }
\shortauthor{T.~Ishikawa, D.~R.~Brumley \& T.~J.~Pedley}

\title{Rheology of a concentrated suspension of spherical squirmers: monolayer in simple shear flow}

\author
 {T.~Ishikawa\aff{1} 
D.~R.~Brumley\aff{2}
T.~J.~Pedley\aff{3}
\corresp{\email{tjp3@damtp.cam.ac.uk}}
}

\affiliation
{
\aff{1}
Department of Finemechanics, Tohoku University, 6-6-01, Aoba, Aramaki, Aoba-ku, Sendai 980-8579, Japan
\aff{2}
School of Mathematics and Statistics, The University of Melbourne, Parkville, Victoria 3010, Australia
\aff{3}
Department of Applied Mathematics and Theoretical Physics, University of Cambridge, Centre for Mathematical Sciences, Wilberforce Road, Cambridge CB3 0WA, UK
}

\maketitle

\begin{abstract} 
A concentrated, vertical monolayer of identical spherical squirmers, which may be bottom-heavy, and which are subjected to a linear shear flow, is modelled computationally by two different methods: Stokesian dynamics, and a lubrication-theory-based method. Inertia is negligible. The aim is to compute the effective shear viscosity and, where possible, the normal stress differences as functions of the areal fraction of spheres $\phi$, the squirming parameter $\beta$ (proportional to the ratio of a squirmer's active stresslet to its swimming speed), the ratio $Sq$ of swimming speed to a typical speed of the shear flow, the bottom-heaviness parameter $G_{bh}$, the angle $\alpha$ that the shear flow makes with the horizontal, and two parameters that define the repulsive force that is required computationally to prevent the squirmers from overlapping when their distance apart is less than a critical value $\epsilon a$, where $\epsilon$ is very small and $a$ is the sphere radius. The Stokesian dynamics method allows the rheological quantities to be computed for values of $\phi$ up to $0.75$; the lubrication-theory method can be used for $\phi> 0.5$. For non-bottom-heavy squirmers, which are unaffected by gravity, the effective shear viscosity is found to increase more rapidly with $\phi$ than for inert spheres, whether the squirmers are pullers ($\beta > 0$) or pushers ($\beta < 0$); it also varies with $\beta$, though not by very much, especially for pushers, and increases with $Sq$. However, for bottom-heavy squirmers the behaviour for pullers and pushers as $G_{bh}$ and $\alpha$ are varied is very different, since the viscosity can fall even below that of the suspending fluid for pushers at high $G_{bh}$; the mechanism for this is not the same as observed in suspensions of bacteria, because the rod-like nature of bacterial cells is believed to dominate in that case. The normal stress differences, which are small for inert spheres, can become very large for bottom-heavy squirmers, increasing with the vigour of the squirming, $\beta$, and varying dramatically as the orientation $\alpha$ of the flow is varied from 0 to $\pi/2$. A major finding of this work is that, despite very different assumptions, the two methods of computation give overlapping results for viscosity as a function of $\phi$ in the range $0.5 < \phi < 0.75$. This suggests that lubrication theory, based on near-field interactions alone, contains most of the relevant physics, and that taking account of interactions with more distant particles than the nearest is not essential to describe the dominant physics. 
\end{abstract}


\section{Introduction}\label{sec:1}

Active matter has become a popular area of research in recent years, among both fluid dynamicists and soft matter physicists. The canonical examples of fluid active matter are suspensions of swimming micro-organisms, which may exhibit fascinating phenomena, ranging from steady, regular patterns, as in bioconvection \citep{Wager:1911,Platt1961,Childress1975,Kessler1986,Pedley1992} among others, to random coherent structures, sometimes referred to as bacterial turbulence \citep{Dombrowski:2004}, with many variants in between. 

There have been few experimental studies on the effective viscosity of suspensions of microscopic swimmers. Notable among those few are the measurements of \cite{Rafai2010} on suspensions of motile algae (\emph{Chlamydomonas reinhardtii}), which are close to spherical, are bottom-heavy, and \emph{pull} themselves through the fluid (equivalent in the squirmer model to $\beta > 0$; see Eq.~\eqref{sq_bc} below). The effective viscosity in a horizontal shear flow was found to increase with volume fraction much more dramatically than for a suspension of dead cells. On the other hand, \cite{Sokolov2009} measured the effective viscosity of a suspension of bacteria (pushers: $\beta < 0$), and found a significant decrease in shear viscosity with swimming speed. The latter finding was reinforced by \cite{Lopez:2015}, using a Couette viscometer in which the flow is essentially simple shear. They found that the presence of pusher cells (\emph{E. coli} bacteria) caused the effective viscosity to fall below that of the ambient fluid at sufficiently low values of the shear rate $\gamma$, and even to approach zero as the cell volume fraction was increased. The mechanism for this phenomenon was first set out by \cite{Hatwalne2004} and depends on the cells being elongated (prolate spheroids or rods): in the absence of swimming, such rods describe Jeffery orbits and align preferentially with the directions of extensional strain rate. The contractile nature of the flow generated by a puller opposes this extension and therefore the effective viscosity is increased, but the extensile nature of pusher-generated flow enhances it. The mechanism is operative for dilute suspensions; it is very clearly explained in the review by \cite{Saintillan:2018}. However, the same mechanism does not apply to a spherical organism, which does not have a preferred orientation in simple shear flow.

Observations of phenomena such as those described above have stimulated an equally large range of theoretical models to see if the observations can be explained by physical processes alone, without requiring an understanding of biological (or chemical) signalling or intracellular processes. Continuum models have been very successful for dilute suspensions, in which the cells interact with their environment but not with each other: bioconvection results from either a gravitational instability when the upswimming of dense cells leads to a gravitationally unstable density profile, or a gyrotactic instability in which the cell's non-uniform density or geometric asymmetry causes them to be reoriented in a shear flow \citep{Childress1975,Kessler1986,Pedley1992,Roberts2002}. Even when gravity is unimportant the stresses applied by the cell's swimming motions (a swimmer acts as a force dipole or stresslet) lead to instability and random bulk motions \citep{Pedley1990,Simha2002,Saintillan2008}.

The rheology of a suspension in an incompressible fluid is represented by the bulk stress tensor ${\bm \Sigma}$, which needs to be calculated or measured in terms of the (changing) configuration of the particles in the suspension and of the velocity field. Rational scientific analysis of the rheology of non-dilute suspensions began with the work of \cite{Batchelor1970}, who showed that ${\bm \Sigma}$, for force-free particles in a (quasi-)steady linear flow with strain-rate tensor $\mathbf{E}$, could be expressed as
\begin{equation}
    {\bm \Sigma} = -P\mathbf{I} + 2\eta_0\mathbf{E} + {\bm \Sigma}^{(p)}, \label{bulk_stress_tensor}
\end{equation}
where the first term is the isotropic part of the stress, $P$ being the effective pressure and the second term is the Newtonian viscous stress ($\eta_0$ being the fluid viscosity). The third term is the particle stress tensor, defined as the average over all spheres of the stresslet for a single particle:
\begin {equation}
\mathbf{S} = \int_{A_p} [\frac{1}{2}\left\{({\bm \sigma} \cdot {\bm n}){\bm x} + {\bm x}({\bm \sigma} \cdot {\bm n})\right\} - \frac{1}{3}{\bm x} \cdot {\bm \sigma} \cdot {\bm n} \mathbf{I } - \eta(\bm{un} + \bm{nu})] dA, \label{Stress_tensor_eqn}
\end{equation}
where ${\bm \sigma}$ is the stress tensor and $\bm{u}$ is the velocity. $A_p$ is the surface of the particle with outward normal $\bm{n}$. An effective viscosity can then be defined for each off-diagonal component of the (ensemble) average stress tensor by dividing it by the corresponding component of the bulk rate of strain tensor, which shows that in general the effective viscosity is itself non-isotropic. Moreover it will in general also depend on the mean velocity field, which means the suspension is non-Newtonian. Further non-Newtonian effects that are of practical importance and are often calculated are the differences between the diagonal stress components (`normal stress differences').

There is a considerable literature on the computation of the effective viscosity and normal stress differences in suspensions of passive particles, in particular identical, rigid, noncolloidal spheres at low Reynolds number. The first contribution was from \cite{Einstein:1906}, who calculated the first correction, of $\mathcal{O}(\phi)$ where $\phi$ is the particle volume fraction, to the fluid shear viscosity; the suspension was dilute and there was no interaction between particles. Semi-dilute suspensions, in which pairwise hydrodynamic interactions were included, were considered by \cite{BatchelorGreenII}, and they could compute the $\mathcal{O}(\phi^2)$ term, at least in linear flows with no closed streamlines. Higher concentrations in general require substantial computations; great progress has been made using the method of Stokesian dynamics, introduced by Brady and his colleagues, e.g. \citep{Bossis:1984,BradyBossis:1985,BradyBossis:1988,BradyPhillips:1988}; application to the rheology of concentrated suspensions was made by \cite{Sierou:2002}. \cite{SinghNott:2000} applied Stokesian dynamics to a monolayer of rigid spherical particles. An excellent recent review of the rheology of concentrated suspensions is given by \cite{Guazzelli:2018}.

In this paper we wish to consider concentrated suspensions of active particles, in which the flow is dominated by cell-cell interactions. Continuum models fail because there is no agreed way of incorporating cell-cell interactions into the model equations – in particular the particle stress tensor ${\bm \Sigma}^{(p)}$ – even when the interactions are restricted to near-field hydrodynamics together with a repulsive force to prevent overlap of model cells. We seek to understand the rheological properties of an idealised suspension from direct numerical simulations. The cells are modelled as identical steady spherical squirmers \citep{Lighthill:1952,Blake:squirmer,Ishikawa:2006,Pedley2016IMA}: spheres of radius $a$ which swim by means of a prescribed tangential velocity on the surface, 
\begin{equation}
u_{\theta} = \frac{3}{2}V_s\sin{\theta}(1 + \beta\cos{\theta}), \label{sq_bc}
\end{equation}
where $\theta$ is the polar angle from the cell's swimming direction,  $V_s$ is the cell swimming speed and $\beta$ represents the stresslet strength; inertia is negligible. The suspension is taken to be a monolayer, in which the sphere centres and their trajectories are confined to a single plane, which is vertical in cases for which gravity, ${\bm g}$, is important. The monolayer is here taken to be confined to the narrow gap between stress-free planes, spaced a distance $2.1a$ apart. In previous work on semi-dilute suspensions, the monolayer was taken to be embedded in an effectively infinite fluid \citep{IshikawaPedley:2008}; we show that the results of the two models do not differ greatly. The spheres may be bottom-heavy, so that when the swimming direction of a sphere, ${\bm p}$, is not vertical the sphere experiences a gravitational torque ${\bm T}$, where 
 \begin{equation}
{\bm T} = -\rho \upsilon h {\bm p}\times {\bm g}
\end{equation}
and $\upsilon, h$ are the cell volume and the displacement of the centre of mass from the geometric centre; $\rho$ is the average cell density, assumed in this case to be the same as that of the fluid. The monolayer is taken to be driven by a simple shear flow in the same, $x$-$y$, plane: ${\bm U} = (\gamma y,0,0)$, with shear-rate $\gamma$; we will also take 
\begin{equation}
{\bm g} = -g(\sin{\alpha},\cos{\alpha},0), \label{alpha_definition}
\end{equation}
so the flow is horizontal if $\alpha=0$.

We have previously studied the rheology of semi-dilute, three-dimensional, suspensions of squirmers -- volume fraction $\phi\leq 0.1$ -- in which pairwise interactions between the cells were summed simply, neglecting interactions involving more than two cells at a time \citep{Ishikawa:2007rheology}. General pairwise interactions were computed exactly using the boundary element method, supplemented by lubrication theory when cells were very close together. Cells were computationally prevented from overlapping by the inclusion of a repulsive interparticle force
\begin{equation}
{\bm F} = \eta_0 a^2 \gamma F_0 \tau e^{-\tau \epsilon}/(1 - e^{-\tau \epsilon})  \hat{{\bm r}},
\label{repulsive}
\end{equation}
where $\hat{{\bm r}}$ is the unit vector along the line of centres and $\epsilon$  is the minimum dimensionless spacing between two cells \citep{BradyBossis:1985}. \cite{Ishikawa:2007rheology} took $\tau$ equal to $1000$ and $F_0$ was varied. It was found that the swimming activity made very little difference to the effective shear viscosity when the spheres were non-bottom-heavy, but the suspension showed significant non-Newtonian behaviour, such as anisotropic effective shear viscosity and normal stress differences, when the cells were bottom-heavy. Moreover, horizontal and vertical shear flows ($\alpha = 0$ or $\pi /2$) gave very different results.
   
Here we use two different methods of simulation for very concentrated monolayer suspensions, with areal fraction $\phi$ up to 0.8. One is a full numerical simulation using Stokesian dynamics \citep{BradyBossis:1988,BradyPhillips:1988,IshikawaLP:2008,IshikawaPedley:2008}, while in the other we assume that every cell interacts only with its immediate neighbours; in that case the interactions are described using lubrication theory alone. This model was recently used to investigate the stability of a regular array of bottom-heavy squirmers, swimming upwards in the absence of an imposed shear flow \citep{BrumleyPedley:2019}. The point of this is to see if lubrication theory alone is good enough to account for all the interesting rheological behaviour of a concentrated suspension, or if some effect of more distant particles is necessary, in the case of squirmers. As well as illuminating the physics, this might make related computations significantly cheaper than the full Stokesian dynamics. For inert and force-free spheres, \cite{Leshansky:2005} reported in a footnote that suppressing all far field interactions made less than $5\%$ difference to the quantities they were computing.

These methods are explained in more detail in the first parts of the next two sections, and their respective results are presented and discussed in the second parts. The findings are compared in the final section, with further discussion and an outline of intended future work.

\section{Rheological properties calculated by Stokesian dynamics}\label{sec:2}

\subsection{Problem settings and numerical method}

\begin{figure}
\begin{center}
\centerline{\includegraphics[scale=0.27]{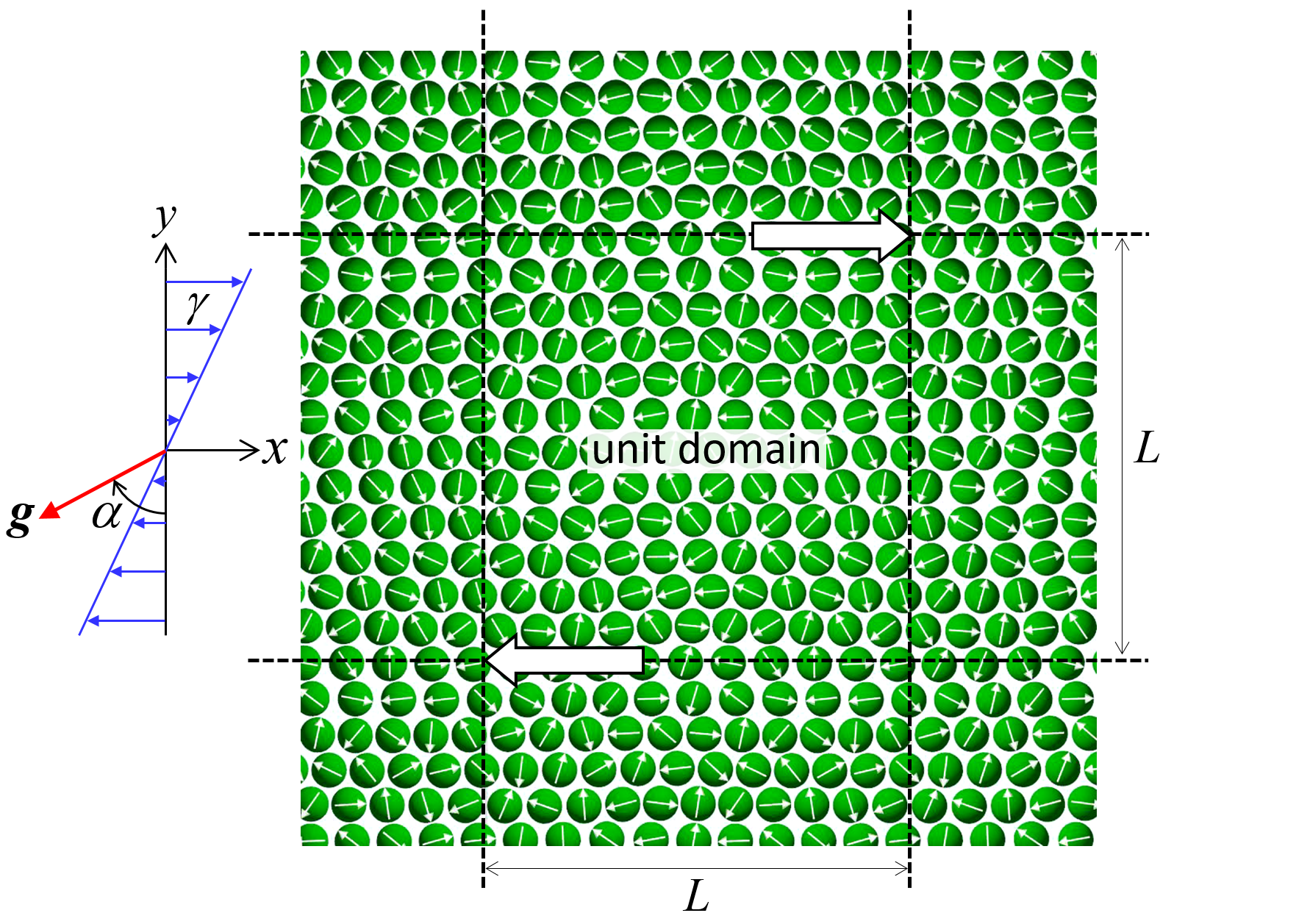}}
\caption{Problem setting of the simulation. An infinitely periodic monolayer suspension of squirmers in a thin fluid film is sheared in the $x-y$ plane. The unit domain is a square with side length $L$, and contains 128 squirmers.
The film is assumed to be flat with thickness $L_z = 2.1a$, and has two stress free surfaces. $\alpha$ is the angle of gravitational acceleration $\bm{g}$ taken from the $-y$ axis.
}
\label{fig1}
\end{center}
\end{figure}

In this study, we consider a monolayer suspension of squirmers in a thin fluid film. The film is assumed as infinitely periodic, and the two surfaces are assumed to be flat and stress free. The film thickness $L_z$ is set as $L_z = 2.1a$ throughout the study.
The reason why we consider a monolayer suspension, instead of a 3-dimensional suspension, is because it allows us to handle a larger system size with a limited number of particles. Moreover, a film suspension can in principle be generated experimentally by using a soap film. As shown in Fig.~\ref{fig1}, in-plane shear flow is induced in the monolayer suspension of squirmers; the $x$-axis is taken in the flow direction, the $y$-axis is taken in the velocity gradient direction, and the $z$-axis in taken perpendicular to the film plane. The background shear flow can be expressed as $( u_x, u_y, u_z) = ( {\gamma}y, 0, 0)$.
Gravity also acts in the $x-y$ plane, and $\alpha$ is the angle the gravitational acceleration ${\bm g}$ makes with the $-y$ axis, as given by Eq.~\eqref{alpha_definition} and shown in the figure.
We assume that body centres and orientation vectors of squirmers are placed in the centre-plane of the fluid film. Thus, due to the symmetry of the problem, the squirmers remain in the same centre plane all the time.

The hydrodynamic interactions among squirmers in an infinite periodic monolayer suspension are calculated in the same manner as in our former studies \citep{IshikawaPedley:2008,IshikawaLP:2008}, so we explain the methodology only briefly here. The Stokesian dynamics method \citep{BradyBossis:1988,BradyPhillips:1988} is employed. The hydrodynamic interactions among an infinite suspension of particles are computed by the Ewald summation technique. By exploiting the Stokesian dynamics method, the force ${\bm F}$, torque ${\bm T}$ and stresslet $\mbox{\bf{S}}$ of squirmers are given by:
\begin{eqnarray}
\left( \begin{array}{c}
\! {\bm F} \! \\
\! {\bm T} \! \\
\! \mbox{\bf{S}} \!
\end{array} \right) 
&=&
\left[ \mbox{\bf{R}}^{far} - \mbox{\bf{R}}_{2B}^{far} + \mbox{\bf{R}}_{2B}^{near} \right]
\left( \begin{array}{c}
 {\bm U} - \langle {\bm u} \rangle \\
 {\bm \Omega} - \langle {\bm \omega} \rangle \\
- \langle \mbox{\bf{E}} \rangle
\end{array} \right) \\
&~& +
 \left[ \mbox{\bf{R}}^{far}  - \mbox{\bf{R}}_{2B}^{far} \right]
\left( \begin{array}{c}
\! -\frac{2}{3}B_1 {\bm p} + \mbox{\bf{Q}}_{sq}  \\
0 \\
- \frac{1}{5}B_2 \left( 3 \bm{pp} - \mbox{\bf{I}} \right) 
\end{array} \right)
\left( \begin{array}{c}
\! {\bm F}_{sq}^{near} \! \! \\
\! {\bm T}_{sq}^{near} \! \! \\
\! \mbox{\bf{S}}_{sq}^{near} \! \! 
\end{array} \right)  ,
\nonumber
\label{mat.4}
\end{eqnarray}
where $\mbox{\bf{R}}$ is the resistance matrix, ${\bm U}$ and ${\bm \Omega}$ are the translational and rotational velocities of a squirmer, $\langle {\bm u} \rangle$ and $\langle {\bm \omega} \rangle$ are the translational and rotational velocity of the bulk suspension, and $\langle \mbox{\bf{E}} \rangle$ is the rate of strain tensor of the bulk suspension.
${\bf Q}_{sq}$ is the irreducible quadrupole providing additional accuracy, which is approximated by its mean-field value (cf. \citep{BradyPhillips:1988}).
Index \emph{far} or \emph{near} indicates far- or near-field interaction, and 2\emph{B} or \emph{Sq} indicates interaction between two inert spheres or two squirmers, respectively.

The computational region is a square of side $L$, and a suspension of infinite extent is modelled with periodic boundary conditions in the $x$ and $y$ directions. In order to express the stress free surfaces of the fluid film, the monolayer is periodically replicated also in the $z$-direction with $2.1a$ intervals.
For the simple shear flow in the $x$, $y$-plane, the periodic conditions in the $x$ and $z$-directions are straightforward. In the $y$-direction, the periodicity requires a translation in the $x$-direction of the spheres at $y=L$, relative to those at $y=0$, by an amount $L \gamma t$ in order to preserve the bulk linear shear flow, where $t$ is time. 
The number of squirmers $N$ in a unit domain is set as $N = 128$. The areal fraction of squirmers is $\phi$ (not the volume fraction $c$), and it is varied in the range $0.1 \le \phi \le 0.75$. 
Parameters in Eq.~\eqref{repulsive} for the repulsive force are set as $F_0 = 10$ and $\tau = 100$.
The initial configuration of the suspended squirmers is generated by first arranging them in a regular array and then applying small random displacements and random orientations.
The dynamic motions are calculated afterwards by the fourth-order Adams-Bashforth time marching scheme. The computational time step is set as $\Delta t =  5 \times 10^{-4}/\gamma$.

The apparent shear viscosity $\eta$ of the film suspension can be calculated from the suspension averaged stresslet $\langle \mbox{\bf{S}} \rangle$. The excess apparent viscosity is given by
\begin{equation}
\frac{\eta_{xy} - \eta_0}{\eta_0} = c \frac{3}{4 \pi a^3 \eta_0 \gamma}
\langle S_{xy} \rangle
=
\phi \frac{1}{\pi a^2 L_z \eta_0 \gamma}
\langle S_{xy} \rangle
\label{excessvis}
\end{equation}
where $c$ is the volume fraction. The areal fraction $\phi$ and $c$ satisfy the relation $c = 4 a \phi/(3L_z)$, where $L_z$ ($= 2.1a$) is the film thickness.
When squirmers are torque-free, the stresslet is symmetric and $\eta_{xy} = \eta_{yx} = \eta$. When squirmers are bottom-heavy, on the other hand, the stresslet becomes asymmetric and $\eta_{xy} \neq \eta_{yx}$.
The dimensionless first and second normal stress differences are defined as
\begin{equation}
N_1 = \phi \frac{1}{\pi a^2 L_z \eta_0 \gamma}
\left( \langle S_{xx} \rangle - \langle S_{yy} \rangle \right)
,~~~
N_2 = \phi \frac{1}{\pi a^2 L_z \eta_0 \gamma}
\left( \langle S_{yy} \rangle - \langle S_{zz} \rangle \right).
\label{normal}
\end{equation}
In order to obtain suspension-averaged properties, stresslet values are averaged over all particles during the time interval $15 \le t \gamma \le 30$, given that the ensemble values reach the steady state.

We introduce a dimensionless number $Sq$, which is the swimming speed, scaled with a typical velocity in the shear flow, i.e.
\begin{equation}
Sq = V_s/{a\gamma}.
\end{equation}
In simulation cases with $Sq > 1$, parameters are modified as $F_0 = 10Sq$, $\Delta t \gamma = 5 \times 10^{-4} / Sq$, to prevent overlapping of particles. Stresslet values are averaged during the longer time interval $75/Sq \le t \gamma \le 150/Sq$, to obtain the steady state values.

\subsection{Results for non-bottom-heavy squirmers}

\begin{figure}
\begin{center}
\centerline{\includegraphics[scale=0.2]{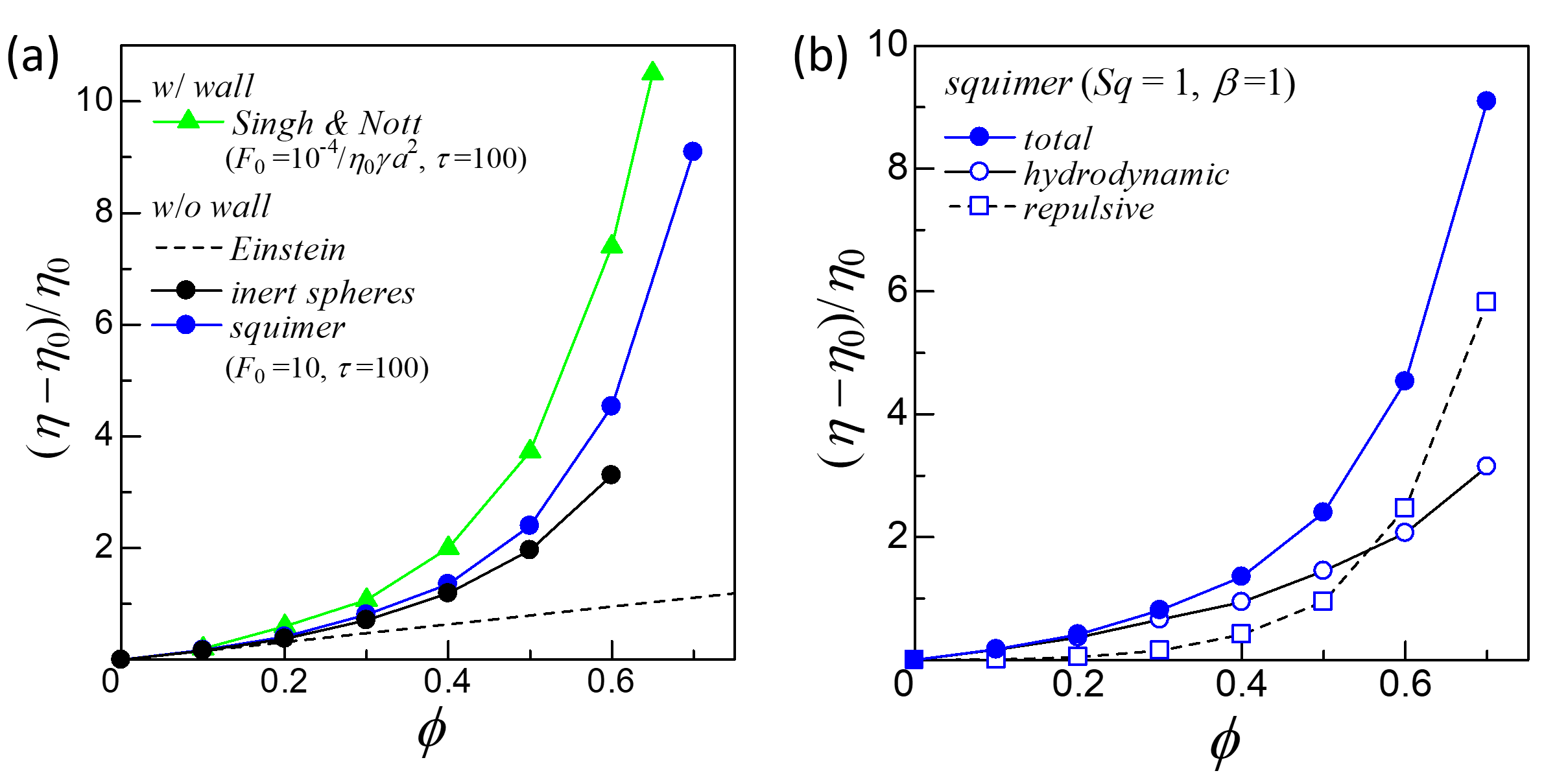}}
\caption{Excess apparent viscosity as a function of areal fraction $\phi$. (a) Present results of inert spheres and squirmers with $Sq = 1$ and $\beta = 1$. The results of \cite{SinghNott:2000} and Einstein's equation of $(\eta - \eta_0)/\eta_0 = 2.5c = (10/6.2) \phi$ are also plotted for comparison. (b) Present results of squirmers with $Sq = 1$ and $\beta = 1$ decomposed into the hydrodynamic contribution and the repulsive contribution.}
\label{fig2}
\end{center}
\end{figure}

We first calculated the apparent viscosity $\eta$ of the suspension, for both inert and squirming spheres. The results are plotted as a function of areal fraction $\phi$ in Fig.~\ref{fig2}(a). We see that $\eta$ of a suspension of squirmers with $Sq = 1$ and $\beta = +1$ (i.e. pullers) increases rapidly with $\phi$. By comparing with the results for inert spheres in the present study, it is found that the motility of a squirmer considerably increases the viscosity. In the case of inert spheres, layers of particles are formed along the flow direction, and near-field interactions between particles are reduced. In the case of squirmers, on the other hand, such layers are destroyed by the motility of the squirmers, which move around irregularly, leading to frequent near-field interactions. Near-field interactions generate large lubrication forces, which result in large particle stresslets as well as the large apparent viscosity.

In Fig. \ref{fig2}, Einstein's equation in the dilute limit as well as the former numerical results of \cite{SinghNott:2000}, for inert spheres, are also plotted for comparison.
We see that all results agree well with Einstein's equation in the dilute regime.
The main differences between the present study and \cite{SinghNott:2000} are the boundary condition and the repulsive interparticle force. In \cite{SinghNott:2000} a monolayer suspension is sheared between two walls, while in the present simulation the shear is generated without a wall boundary. The coefficients of repulsive force given by Eq.~\eqref{repulsive} are $\eta_0 a^2 \gamma F_0 = 10^{-4} [N], \tau = 100 [-]$ in \cite{SinghNott:2000}, while $F_0 = 10 [-], \tau = 100 [-]$ in the present simulation. We see that the apparent viscosity of a suspension of inert particles reported by \cite{SinghNott:2000} is larger than that in the present study. The discrepancy may come from the differences in the boundary condition and the repulsive interparticle force. By introducing the wall boundary, motions of spheres in the $y$-direction are restricted. The restricted motions may enhance the near-field interactions between particles, which results in the larger apparent viscosity. Moreover, the repulsive force may be larger in the present study than in \cite{SinghNott:2000} (it is hard to find the value of the dimensionless coefficient analogous to $F_0$ in that paper). This is because squirmers easily overlap if the repulsive forces are not strong enough, given that lubrication flow between two squirming surfaces cannot mathematically prevent the overlapping. The strong repulsive force may reduce the apparent viscosity, which may be another reason for the discrepancy.

The stresslet can be decomposed into the hydrodynamic contribution and the repulsive contribution, as shown in Fig. \ref{fig2}(b). The repulsive contribution to the particle bulk stress can be calculated as \citep{Batchelor1977}
\begin{equation}
- \frac{1}{V} \sum_{i=2}^N \sum_{j<i} \vec{r}^{ij} \vec{F}^{ij},
\label{stress.4}
\end{equation}
where $V$ is a unit volume, $\vec{r}^{ij}$ is the centre-centre separation of squirmers $i$ and $j$, and $\vec{F}^{ij}$ is their pairwise interparticle force given by Eq.~\eqref{repulsive}. We see that the hydrodynamic contribution is dominant in the low $\phi$ regime. When $\phi \ge 0.6$, on the other hand, the repulsive contribution becomes larger than the hydrodynamic contribution, which is the reason why the apparent viscosity rapidly increases in the high $\phi$ regime.

\begin{figure}
\begin{center}
\centerline{\includegraphics[scale=0.2]{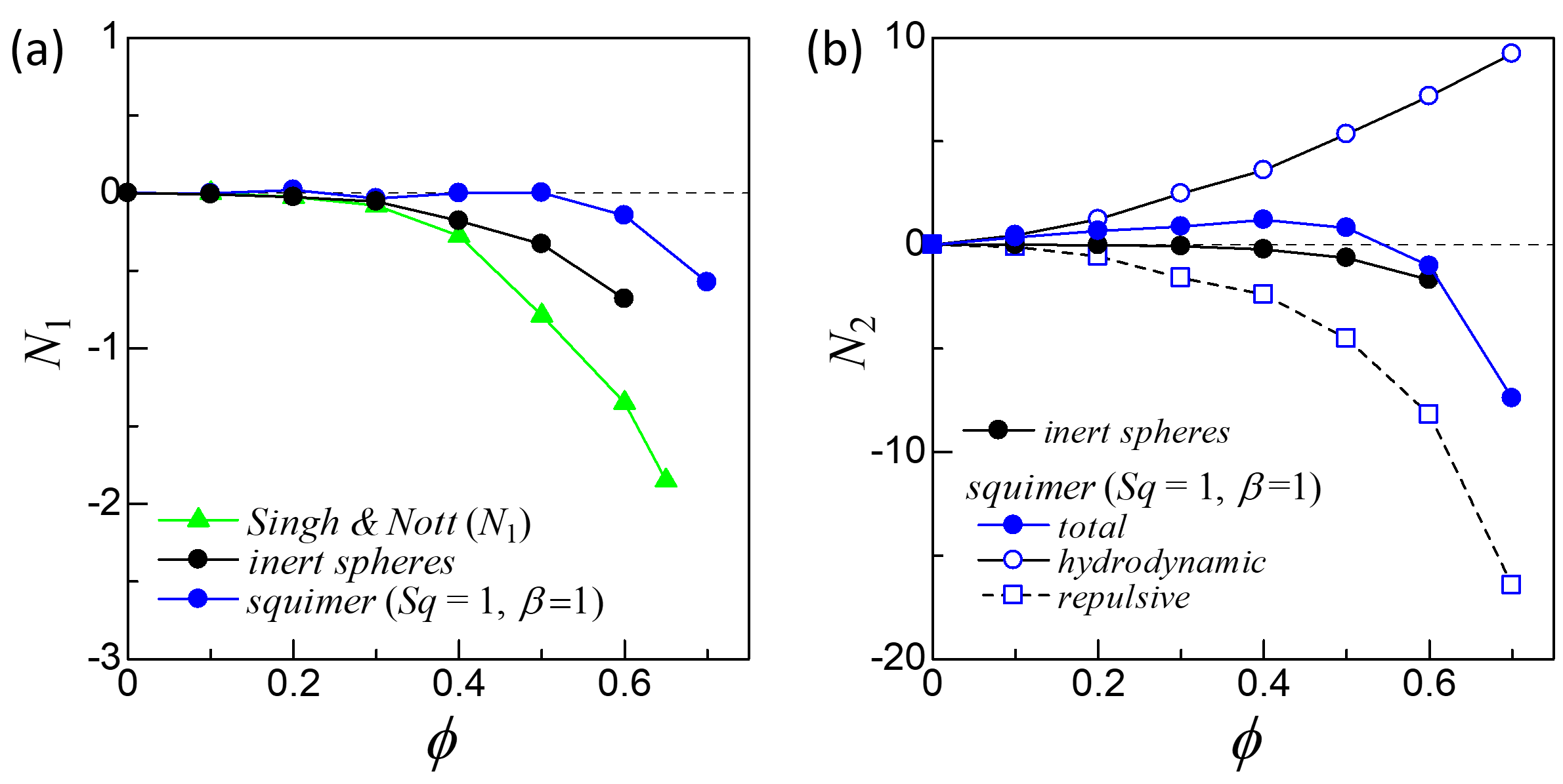}}
\caption{Normal stress differences as a function of areal fraction $\phi$.
(a) $N_1$ in suspensions of inert spheres and squirmers with $Sq = 1$ and $\beta = 1$. The results of \cite{SinghNott:2000} are also plotted for comparison. 
(b) $N_2$ in suspensions of inert spheres and squirmers with $Sq = 1$ and $\beta = 1$. The results of squirmers are decomposed into the hydrodynamic contribution and the repulsive contribution.}
\label{fig3}
\end{center}
\end{figure}

First and second normal stresses also appear in the suspension of inert spheres and squirmers, as shown in Fig. \ref{fig3}. $N_1$ is negative in sign, so particles are basically compressed in the flow direction. $|N_1|$ increases as $\phi$ is increased, similar to the apparent viscosity. The value of $|N_1|$ in a suspension of inert particles, as reported by \cite{SinghNott:2000}, is larger than in the present study. The discrepancy may again come from the differences in the boundary condition and the repulsive interparticle force.

$N_2$ in the suspension of inert spheres is also negative in sign, so the particle stresses satisfy $\Sigma^{(p)}_{xx} < \Sigma^{(p)}_{yy} < \Sigma^{(p)}_{zz}$. 
 In the case of squirmers, the sign of $N_2$ changes at around $\phi = 0.55$. The hydrodynamic contribution to $N_2$ is positive and steadily increases with $\phi$, while the repulsive contribution to $N_2$ is negative and rapidly increases with $\phi$. Since the repulsive contribution overwhelms the hydrodynamic contribution when $\phi \ge 0.6$, $N_2$ becomes negative in the large $\phi$ regime.

\begin{figure}
\begin{center}
\centerline{\includegraphics[scale=0.2]{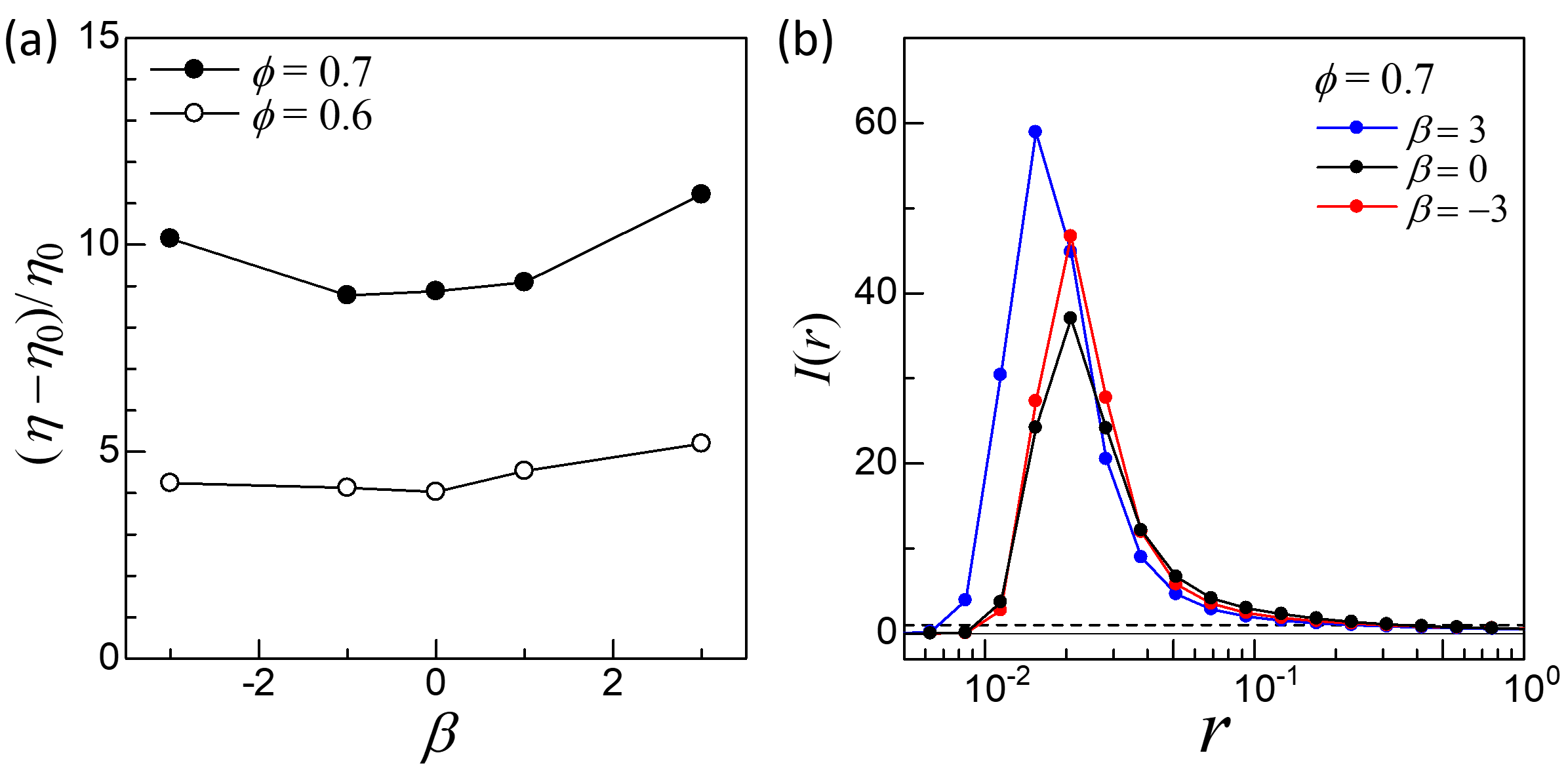}}
\caption{Effect of the swimming mode $\beta$ on the viscosity ($Sq = 1$).
(a) Excess apparent viscosity with $\phi = 0.6$ and 0.7.
(b) Normalised probability density function distribution of squirmers with $\beta = 3$, $0$ and $-3$ ($\phi = 0.7$).
}
\label{fig4}
\end{center}
\end{figure}

The difference between pushers and pullers can be investigated by changing the swimming mode parameter $\beta$, which is positive for pullers and negative for pushers. Figure \ref{fig4}(a) shows the effective viscosity $\eta$ for $Sq=1$ and various values of $\beta$. We see that $\eta$ increases as the absolute value of $\beta$ is increased. This is probably because strong squirming velocity with large $|\beta|$ enhances near-field interactions between squirmers, which induces a strong stresslet. The effect of $\pm \beta$ is not symmetric, but pullers generate larger viscosity than pushers. In order to confirm these tendencies, we calculated the normalised probability density function of squirmers, defined as
\begin{equation}
I(r) = 
\frac{\int_{r=\text{const}} P(r_0|r_0+r) dr}
{2 \pi r \phi} ,
\label{prob.1}
\end{equation}
where $P(r_0|r_0+r)dr$ is the conditional probability that, given that there is a squirmer centred at $r_0$, there is an additional squirmer centred between $r_0 + r$ and $r_0 + r + dr$.
The results are plotted in Fig.~\ref{fig4}(b). We see that $I(r)$ with $\beta = 3$ has a considerably larger value than that with $\beta = 0$ in the small $r$ regime, while $I(r)$ with $\beta = -3$ does not. On the other hand, $I(r)$ with $\beta = -3$ has a larger peak than that with $\beta = 0$.
Thus, near-field interactions between squirmers are increased as $|\beta|$ is increased. The difference between pushers and pullers is obvious in Fig.~\ref{fig4}(b): pullers tend to come closer to each other than pushers. A former study reported that the face-to-face configuration is stable for puller squirmers while unstable for pusher squirmers, though the face-to-face configuration was expressed as a stress-free surface in the paper \citep{Ishikawa2019}. Thus, puller squirmers facing each other can come closer more easily than pushers.

\begin{figure}
\begin{center}
\centerline{\includegraphics[scale=0.2]{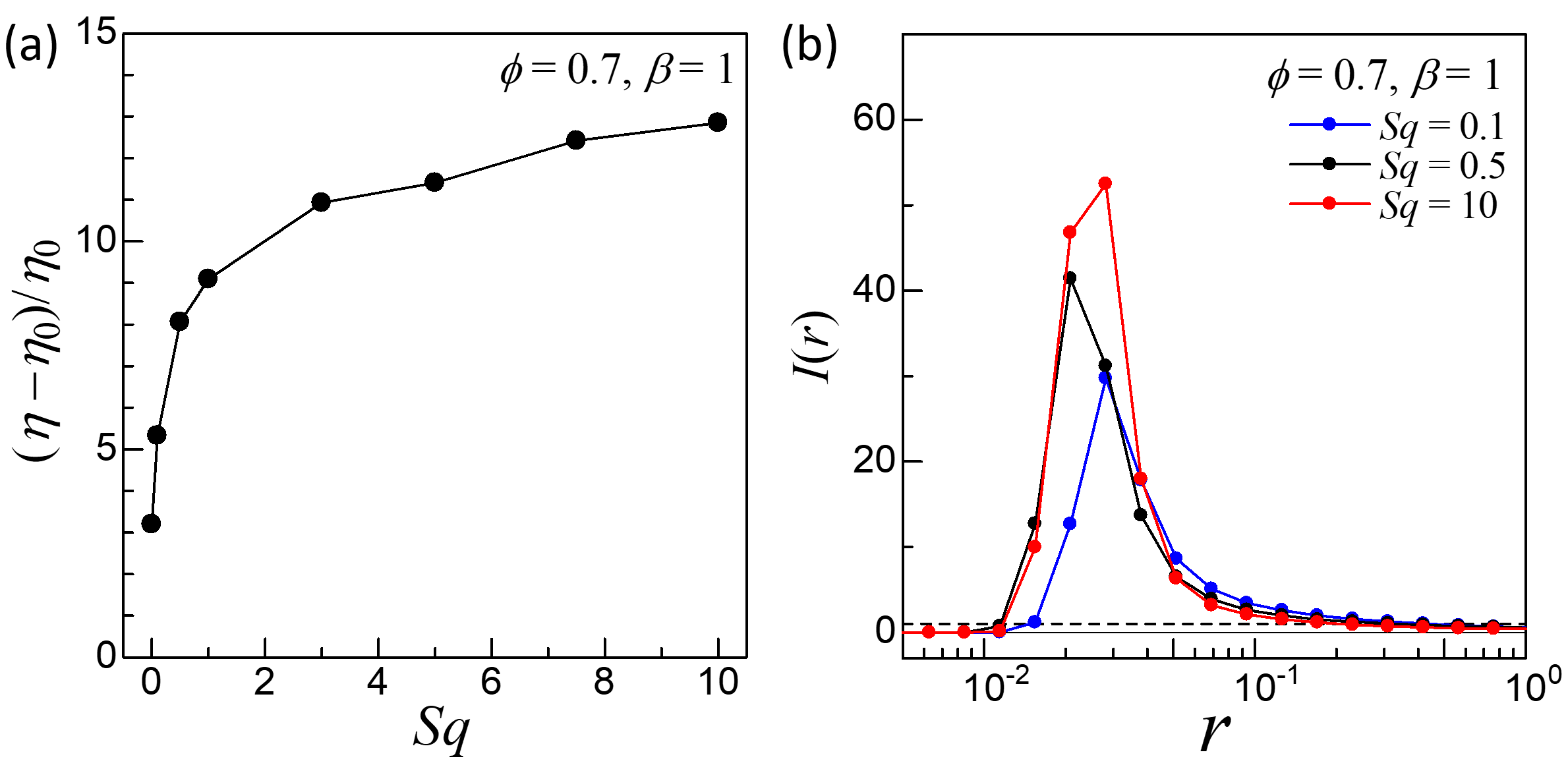}}
\caption{Effect of $Sq$ on the viscosity ($\phi = 0.7$ and $\beta = 1$).
(a) Excess apparent viscosity. (b) Normalised probability density function distribution with $Sq = 0.1, 0.5$ and 5.
}
\label{fig5}
\end{center}
\end{figure}

We also investigated the effect of $Sq$, i.e. the ratio of swimming velocity to the shear velocity. The results are shown in Fig.~\ref{fig5}. We see that the apparent viscosity $\eta$ increases as $Sq$ is increased.
This is because large swimming velocity, relative to the shear velocity, destroys the formation of layers by the particles, and enhances their near-field interactions. This tendency can be confirmed by Fig.~\ref{fig5}(b), in which $I(r)$ with $Sq = 10$ has larger values than with $Sq = 0.1$ and 0.5.
Thus, increase in the apparent viscosity can be understood as coming from the increase of near-field interactions.

\subsection{Results for bottom-heavy squirmers}

When squirmers are bottom-heavy, cells tend to align in the direction opposite to  gravity. To discuss the effect of bottom-heaviness, we introduce a dimensionless number $G_{bh}$, defined as
\begin{equation}
G_{bh}=\frac{4 \pi \rho g a h}{3 \eta_0 V_s} .
\label{Gbh}
\end{equation}
$G_{bh}$ is proportional to the ratio of the time to swim a body length to the time for the axis orientation $\bm{p}$ of a non-swimmer to rotate from horizontal to vertical.
When $G_{bh}$ is sufficiently large, the gravitational torque due to the bottom-heaviness balances the hydrodynamic torque due to the background shear flow, and squirmers tend to align in a particular direction.

\begin{figure}
\begin{center}
\centerline{\includegraphics[scale=0.21]{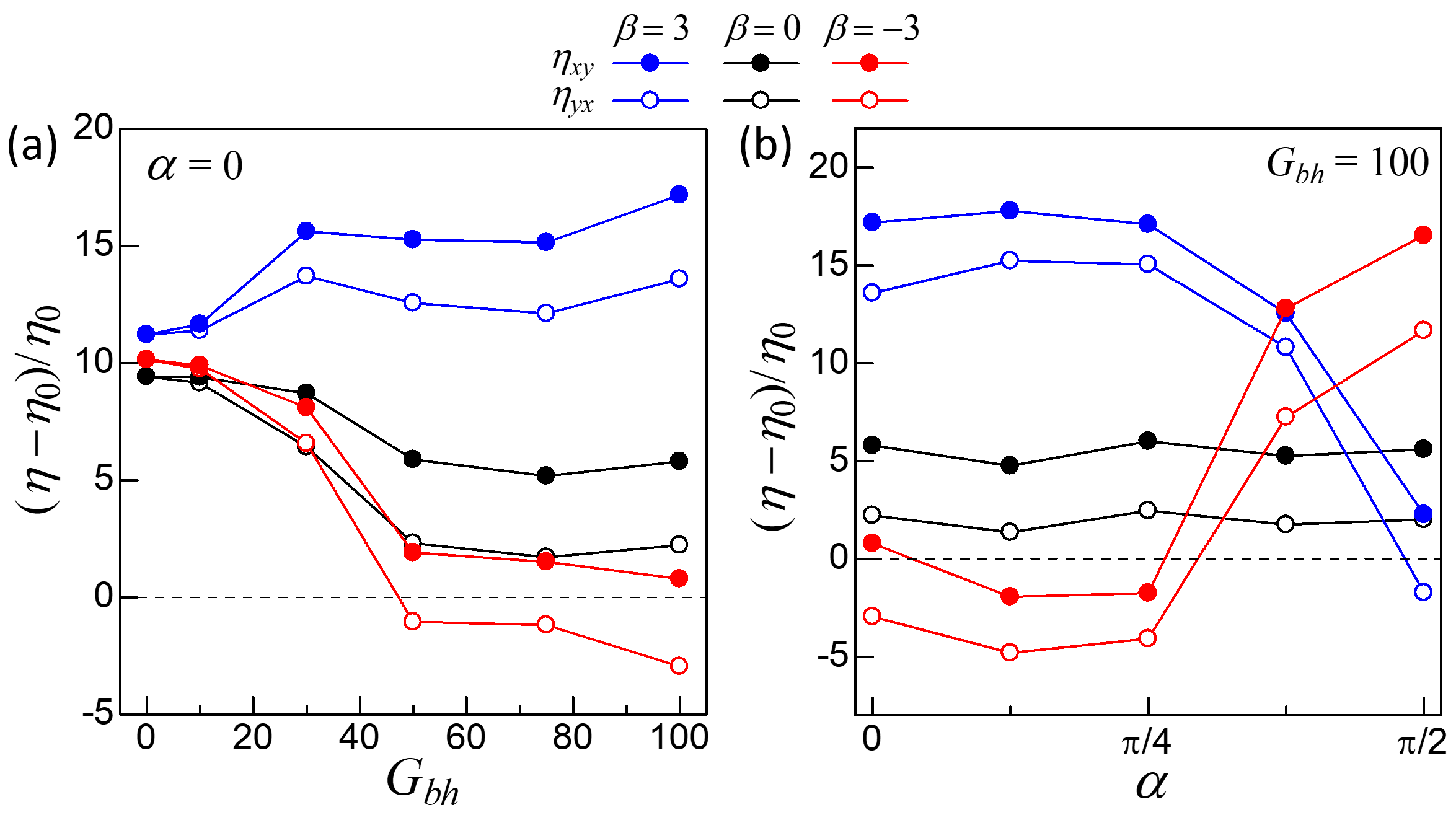}}
\caption{Effect of bottom-heaviness on the excess apparent viscosity ($Sq = 1$, $\phi = 0.7$, $\beta = 3$, $0$ and $-3$).
(a) Effect of $G_{bh}$ ($\alpha = 0$). (b) Effect of the angle of gravity $\alpha$ ($G_{bh} = 100$).
}
\label{fig6}
\end{center}
\end{figure}

The effect of $G_{bh}$ and the gravitational angle $\alpha$ on the apparent viscosity is shown in Fig.~\ref{fig6}. Since an external torque due to the bottom-heaviness is exerted on a squirmer, the partcle stress tensor becomes asymmetric, and the $xy$ and $yx$ components become different. We see that $G_{bh}$ considerably affects the value of $\eta$. For $\beta = 0$ and $-3$, in horizontal flow ($\alpha =0$), $\eta$ is decreased by the bottom-heaviness, and $\eta_{yx}$ with $\beta = -3$ even becomes negative when $G_{bh} \ge 50$. The decrease in the apparent viscosity may be caused by two mechanisms: (a) squirmers with large $G_{bh}$ tend to swim in the same direction, and cell-cell collisions are suppressed; and (b) aligned squirmers induce a net squirming stresslet when $\beta \ne 0$, which directly contributes to $\eta_{xy}$ and $\eta_{yx}$.
The effect of $\alpha$ is also significant: $\eta$ with $\beta = 3$ takes its maximum values around $\alpha = \pi/8$, whereas that with $\beta = -3$ takes its minimum values around $\alpha = \pi/8$. So the tendencies are almost opposite between the pusher and the puller.

\begin{figure}
\begin{center}
\centerline{\includegraphics[scale=0.2]{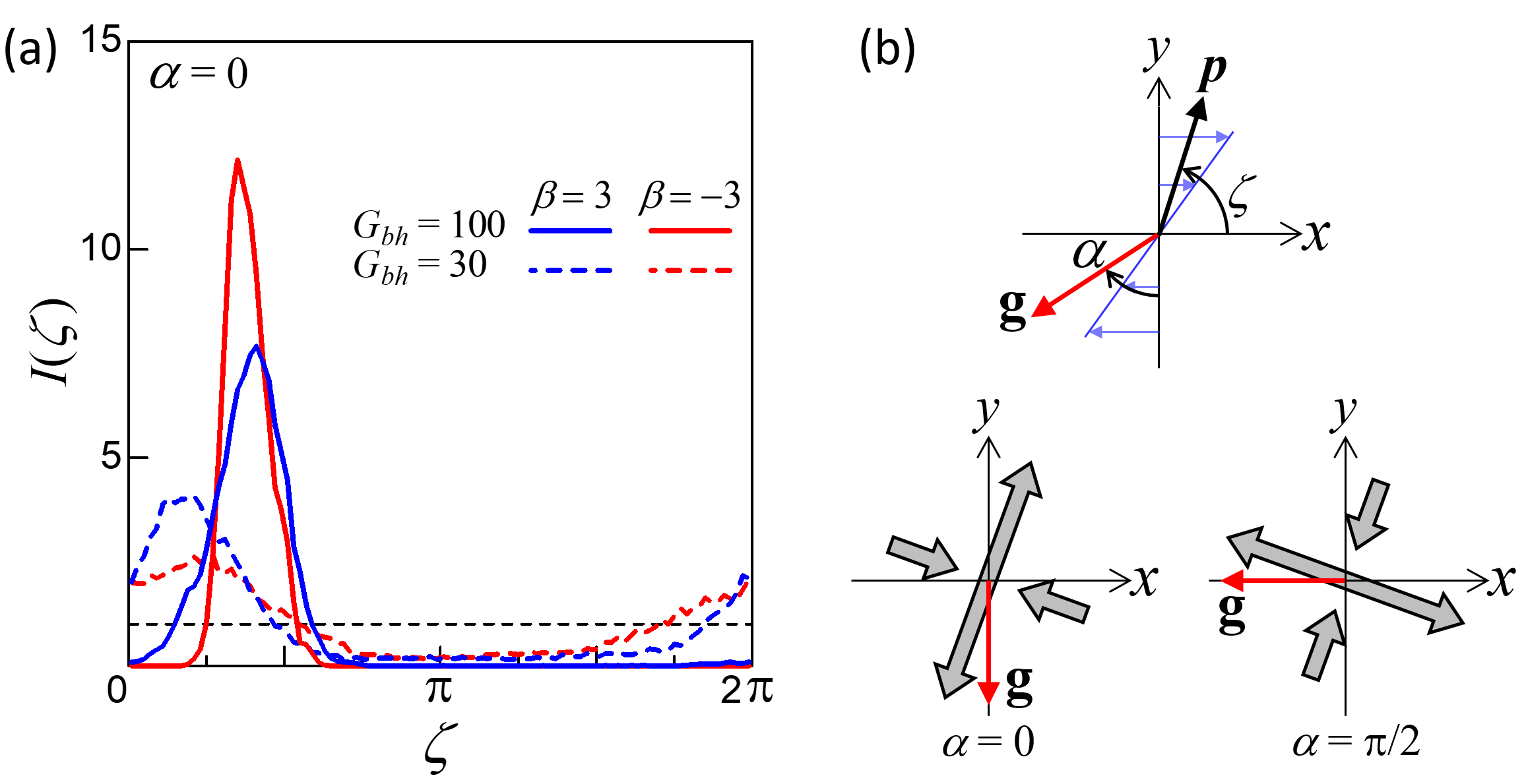}}
\caption{Orientation of bottom-heavy squirmers in shear flow.
(a) Normalised probability density distribution as a function of angle $\zeta$ ($Sq = 1$, $\phi = 0.7$, $\alpha = 0$, $G_{bh} = 30$ and 100, $\beta = 3$ and $-3$).
(b) Definition of angles, and schematics of stresslet directions under different angle of gravity $\alpha$.
}
\label{fig7}
\end{center}
\end{figure}

In order to clarify the mechanism of bottom-heavy effects, we here discuss the orientation of bottom-heavy squirmers in shear flow.
Figure \ref{fig7} shows normalised probability density distribution as a function of angle $\zeta$ that is defined from the $x$-axis in the counter clockwise direction as shown in Fig.~\ref{fig7}(b). When $I(\zeta) = 1$ for all $\zeta$, the orientation distribution is isotropic.
We see that the peak of $I(\zeta)$ is higher, and at a larger value of $\zeta$, in the $G_{bh} = 100$ case than in the $G_{bh} = 30$ case due to the strong bottom-heaviness. Squirmers with $G_{bh} = 100$ are oriented with approximately $\zeta = 0.38\pi$ regardless of $\beta$.
When $\beta = -3$, i.e. pushers, the stresslet with $\alpha = 0$ is directed as in Fig.~\ref{fig7}(b) bottom-left, and the apparent viscosity decreases. If the direction of gravity is rotated to $\alpha = \pi/2$, as in Fig.~\ref{fig7}(b) bottom-right, the direction of the stresslet becomes opposite, and the apparent viscosity increases. These schematics can qualitatively explain the results in Fig. \ref{fig6}. When the squirmers are pullers, the sign of the stresslet is opposite to that shown in Fig.~\ref{fig7}(b). Thus, the apparent viscosity increases with $\alpha = 0$ whereas it decreases with $\alpha = \pi/2$, which is again consistent with Fig. \ref{fig6}. We note that the internal configuration, at $\alpha = 0$, is more regular for $\beta = -3$ than for $\beta = +3$; this is indicated by the higher peak for $\beta = -3$ in Fig.~\ref{fig7}(a). For $\beta = +3$, large numbers of the squirmers are aggregated and almost jam the system.

\begin{figure}
\begin{center}
\centerline{\includegraphics[scale=0.2]{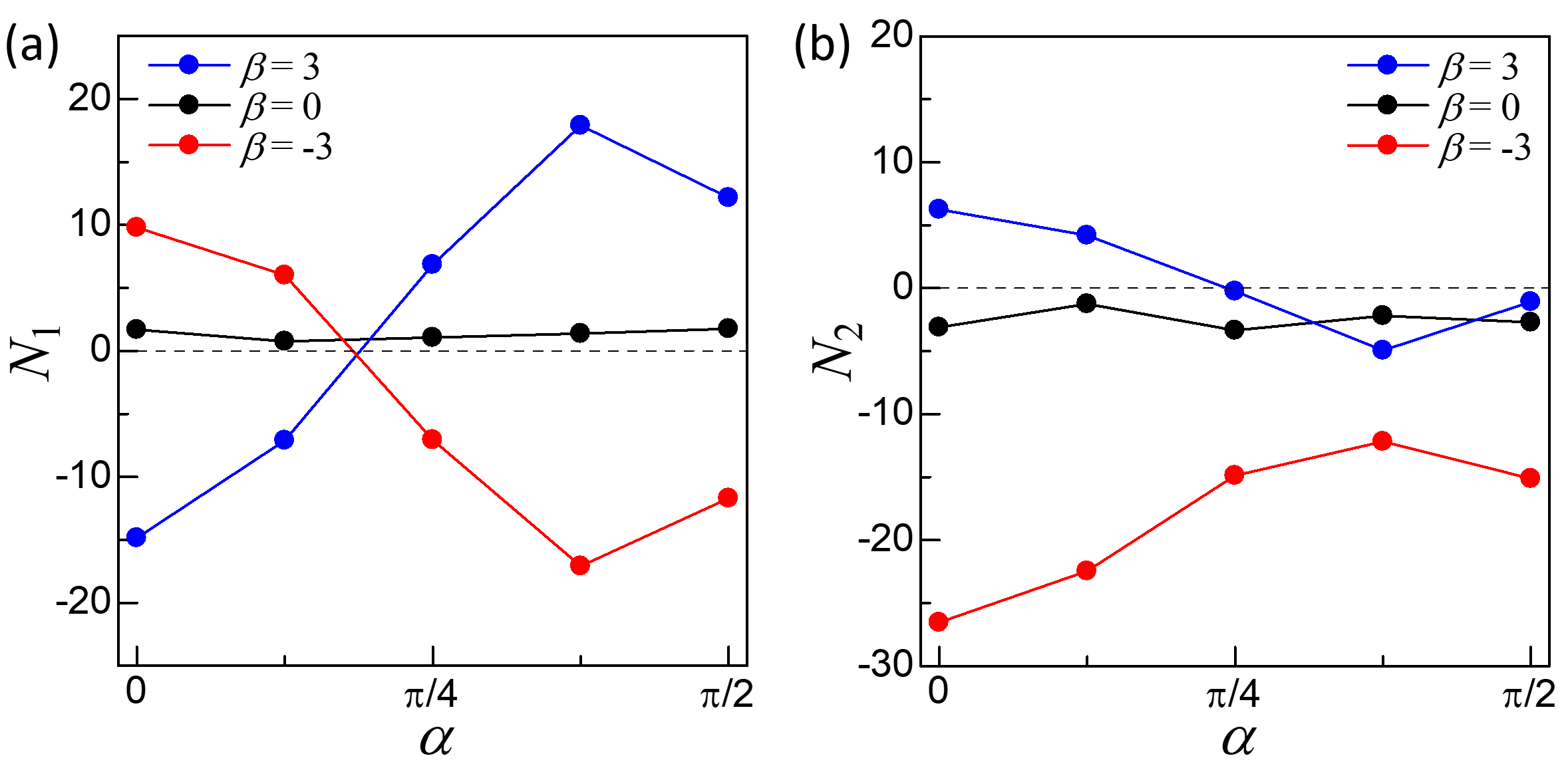}}
\caption{Effect of the angle of gravity $\alpha$ on the normal stress differences ($Sq = 1$, $\phi = 0.7$, $G_{bh} = 100$, $\beta = 3$, $0$ and $-3$).
(a) First normal stress difference. (b) Second normal stress difference.
}
\label{fig8}
\end{center}
\end{figure}

The first and second normal stress differences are affected by the angle of gravity $\alpha$, as shown in Fig.~\ref{fig8}. The first normal stress difference with $\beta = 3$ increases as $\alpha$ is increased, whereas that with $\beta = -3$ decreases.
The second normal stress difference with $\beta = 3$ takes its minimum values around $\alpha = 3\pi/8$, whereas that with $\beta = -3$ takes its maximum values around $\alpha = 3\pi/8$. So the tendencies are again almost opposite between the pusher and the puller. The opposite tendency can be explained by the sign and rotation of the stresslet, as schematically shown in Fig.~\ref{fig7}(b).

\begin{figure}
\begin{center}
\centerline{\includegraphics[scale=0.2]{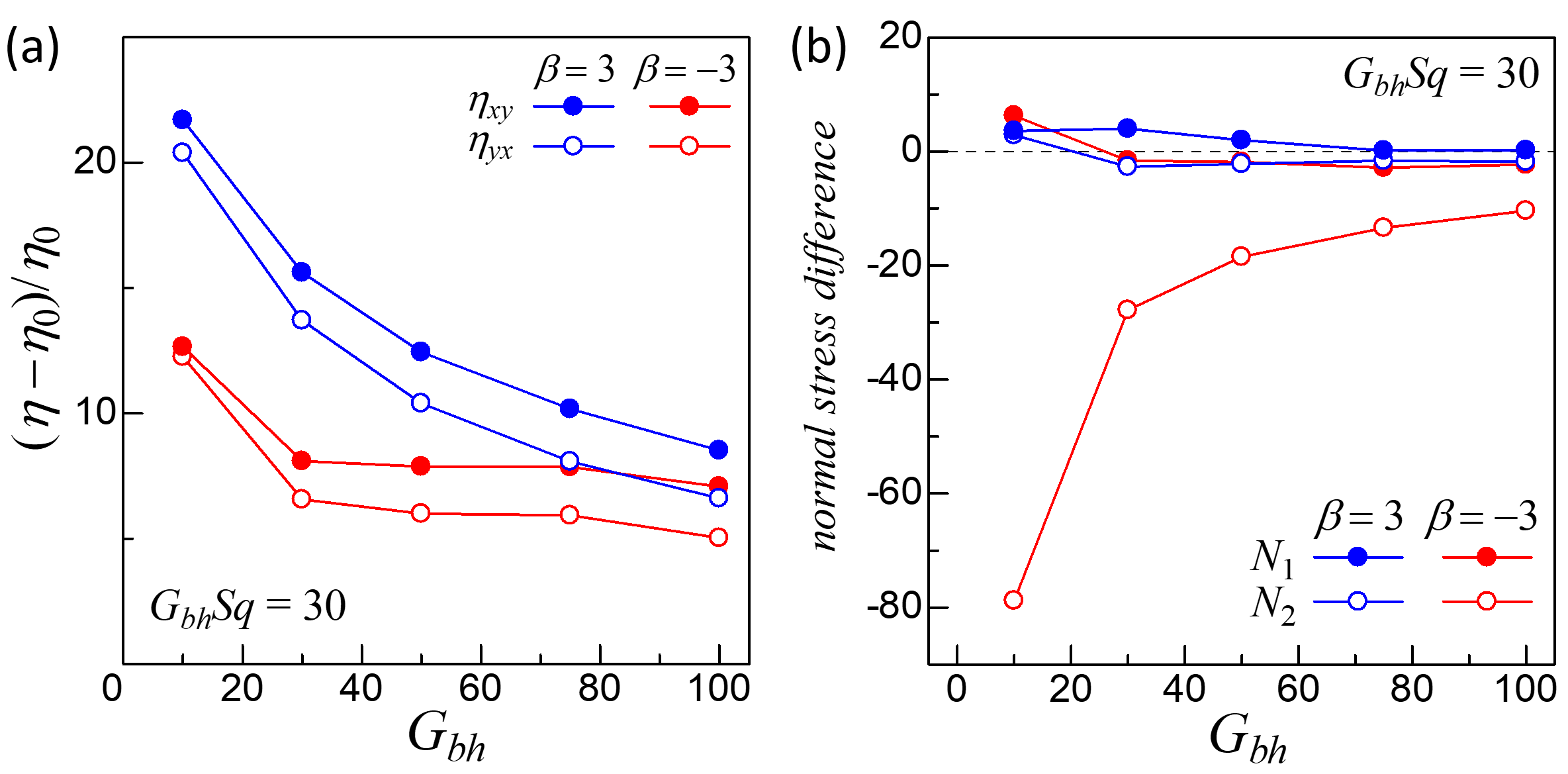}}
\caption{Rheology under the condition of $Sq \cdot G_{bh}$ = 30, in which the orientation angle of a solitary squirmer relative to the gravity is expected to be the same ($\phi = 0.7$, $\beta = \pm 3$, $\alpha = 0$).
(a) Excess apparent viscosity and (b) Normal stress differences.}
\label{fig9}
\end{center}
\end{figure}

Last, we investigate the rheology under constant $Sq \cdot G_{bh}$ conditions. This product represents the ratio of the gravitational torque to the shear torque, independently of the swimming speed $V_s$. Thus, a solitary squirmer under constant $Sq \cdot G_{bh}$ conditions is expected to have the same orientation angle relative to gravity. The results for excess apparent viscosity and normal stress differences under the condition of $Sq \cdot G_{bh} = 30$ are shown in Fig. \ref{fig9}. We see that $\eta$ is considerably increased in the small $G_{bh}$ regime. The effect of swimming is larger than that of the background shear in the small $G_{bh}$ regime. The large swimming velocity enhances near-field interactions between squirmers, which results in the large apparent viscosity. The normal stress difference $N_2$ for $\beta = -3$ is also enhanced in the small $G_{bh}$ regime. This is because the active stresslet, caused by the squirming velocity, plays a dominant role compared to the passive stresslet, caused by inert spheres, in the small $G_{bh}$ regime. Hence, the rheology under constant $Sq \cdot G_{bh}$ conditions is strongly affected by $Sq$ especially when it is large.

\newpage
\section{Shear viscosity calculated by lubrication interactions} \label{sec:3}

\subsection{Problem setting and numerical method}

In this Section, we present a complementary method for calculating the rheological properties of a suspension of steady, spherical squirmers. The methodology for measuring the bulk viscosity is similar to that of a rheometer; a suspension of squirmers situated between flat parallel plates is sheared using Couette flow, and the drag force on the plates is measured. The overall dynamics of the active suspension are solved by summing pairwise lubrication interactions between closely-separated squirmers, along with hydrodynamic forces and torques associated with solitary squirmers in shear flow. This approach utilises elements of previous work \citep{BrumleyPedley:2019}, but with several key advances. Firstly, in the present formulation, an arbitrary areal fraction of squirmers is permitted, so that cells are no longer confined to a hexagonal crystal array. Secondly, the entire suspension will be subject to a background shear flow. The dependence of the rheology on a number of key suspension and external parameters are presented. Despite the key differences between this approach and that of Section~\ref{sec:2}, the results are in good quantitative agreement at sufficiently large areal fractions. \\

\begin{figure}
\begin{center}
\centerline{\includegraphics[width=\textwidth]{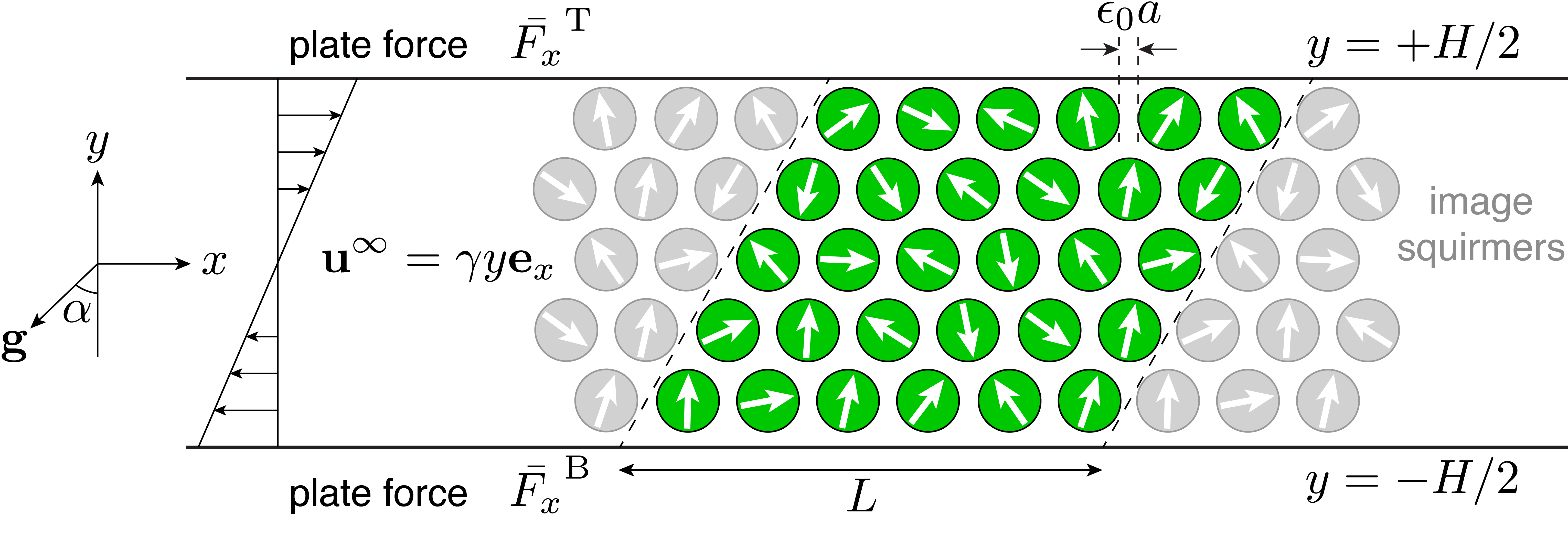}}
\caption{A suspension of $N = n_x n_y = 132$ spherical squirmers, subject to a background shear flow $\mathbf{u}^{\infty} = \gamma y \bm{e}_x$ between two parallel plates. The position $\mathbf{x}_i$, and orientation $\mathbf{e}_i$ of each squirmer is confined to lie in the $x$-$y$ plane. The direction of gravity is $\mathbf{g} = -g(\sin \alpha, \cos \alpha, 0 )$.}
\label{fig10}
\end{center}
\end{figure}

Two infinite no-slip plates situated at $y_1= +H/2$ and $y_2=-H/2$ move with velocities $\gamma y_1 \bm{e}_x$ and $\gamma y_2 \bm{e}_x$, respectively, so that the fluid between the plates is subject to a uniform shear with rate $\gamma$. That is, the fluid velocity is given by $\bm{u} = (u_x, u_y, u_z) = (\gamma y, 0, 0)$. A suspension of $N$ identical squirmers is introduced into the fluid (see Fig.~\ref{fig10}). 

The position and orientation vectors of all squirmers are restricted to lie in the $x$-$y$ plane, so that each squirmer has $2+1$ degrees of freedom. Moreover, the domain is subject to periodic boundary conditions in the $x$-direction, with period $L$ such that the squirmer areal fraction $\phi = N \pi a^2 /LH$, where $N=n_x n_y$. This can also be expressed in the following way:
\begin{equation}
\phi =  \frac{N \pi a^2}{LH} =  \frac{2 \pi n_y}{\sqrt{3} (2+\epsilon_0) \big[(2+\epsilon_0)n_y + \epsilon_0 \big]},
\end{equation}
where $\epsilon_0 a $ is the spacing between adjacent squirmers in the case where they are arranged in a regular hexagonal array (as depicted in Fig.~\ref{fig10}). The total dimensional force $\bm{F}_i$ and torque $\bm{T}_i$ on the $i^{\text{th}}$ squirmer are composed of several contributions, as outlined below. As in \cite{BrumleyPedley:2019}, these will be scaled by $\eta_0 \pi a$ and $\eta_0 \pi a^2$ respectively, so that they have units of velocity. As in Section~\ref{sec:2}, the parameter $\eta_0$ is the viscosity of the solvent around the squirmers. The resistance formulation for the spheres in Stokes flow is given by 
\begin{align}
\left( \begin{array}{c}
\bar{\bm{F}} \\
\bar{\bm{T}} \\
\end{array} \right)^{\text{net}} &=
\bm{R} \cdot
\left( \begin{array}{c}
\bm{V} \\
a \bm{\omega} \\
\end{array} \right) + 
\left( \begin{array}{c}
\bar{\bm{F}}^{\text{sq}} + \bar{\bm{F}}^{\text{rep}} + \bar{\bm{F}}^{\text{prop}} + \bar{\bm{F}}^{\infty} \\
\bar{\bm{T}}^{\text{sq}} + \bar{\bm{T}}^{\text{grav}}  + \bar{\bm{T}}^{\infty} \\
\end{array} \right), \label{lubrication_resistance}
\end{align}
where the resistance matrix is given by $\bm{R} = \bm{R}^{\text{sq-sq}} + \bm{R}^{\text{sq-wall}} + \bm{R}^{\text{drag}}$. The vectors $\bar{\bm{F}} = \bm{F} / (\eta_0 \pi a)$ and $\bar{\bm{T}} = \bm{T} / (\eta_0 \pi a^2)$ are of length $2N$ and $N$ respectively, and are composed of the forces $[F_x^{(i)},F_y^{(i)}]$ and torques $[T_z^{(i)}]$ for all squirmers $i \in \{1,2,\dots, N\}$. The corresponding vectors $\bm{V}$ and $a \bm{\omega}$ contain the linear and angular velocities, respectively, of all squirmers in the suspension. At zero Reynolds number, all squirmers must be force- and torque-free. Equation~\eqref{lubrication_resistance} therefore reduces to 
\begin{align}
\bm{R} \cdot
\left( \begin{array}{c}
\bm{V} \\
a \bm{\omega} \\
\end{array} \right) &= 
- \left( \begin{array}{c}
\bar{\bm{F}}^{\text{sq}} + \bar{\bm{F}}^{\text{rep}} + \bar{\bm{F}}^{\text{prop}} + \bar{\bm{F}}^{\infty} \\
\bar{\bm{T}}^{\text{sq}} + \bar{\bm{T}}^{\text{grav}}  + \bar{\bm{T}}^{\infty} \\
\end{array} \right). \label{lubrication_resistance2}
\end{align}
Explicit hydrodynamic coupling between squirmers is limited to lubrication interactions. For two squirmers $i$ and $j$, with minimum clearance $\epsilon_{ij} = (|\mathbf{x}_i - \mathbf{x}_j|-2a)/a$ (subject to periodic boundary conditions), lubrication interactions occur only if $\epsilon_{ij} < 1$. This value is chosen because $\log \epsilon_{ij}$, the functional dependence of these forces and torques, is equal to zero at that point. These terms therefore increase continuously from zero as squirmers approach one another from afar.
The matrix $\bm{R}^{\text{sq-sq}}$ captures the hydrodynamic forces and torques acting on every pair of spheres sufficiently close to one another, arising from their linear and angular velocities. These expressions, evaluated for no-slip spheres, can be found in \cite{KimAndKarrila:Microhydrodynamics} and \cite{BrumleyPedley:2019}. In a similar fashion, $\bm{R}^{\text{sq-wall}}$ captures the forces and torques due to motion of the spheres in close proximity to the bounding walls. The matrices $\bm{R}^{\text{sq-sq}}$ and $\bm{R}^{\text{sq-wall}}$, composed of blocks for each pair of squirmers, are non-zero only for pairs that are sufficiently close to permit lubrication interactions. The sparsity pattern of these resistance matrices therefore depends on the physical configuration of the system, and must be computed at every time step. Conversely, the final component of the resistance matrix is always diagonal, and is given by 
\begin{align}
\bm{R}^{\text{drag}} = 
\left( \begin{array}{c|c}
-6 \bm{I}_{2N} & \bm{0} \\
\hline
\bm{0} & -8 \bm{I}_{N} \\
\end{array} \right),
\end{align}
where $\bm{I}_{n}$ is the $n \times n$ identity matrix. This captures the drag on a solitary translating and rotating sphere in Stokes flow. Inclusion of this matrix ensures that even widely separated spheres that do not experience lubrication interactions, are subject to solitary Stokes drag.

There are a number of contributions to the forces and torques that are independent of the squirmer velocities. First, there are the contributions which depend on the arrangement of cells with respect to one another. The squirming motions generate contributions for all pairs that are sufficiently close to one another. We refer the reader to \cite{BrumleyPedley:2019} for detailed expressions of these forces and torques, $\bar{\bm{F}}^{\text{sq}}$ and $\bar{\bm{T}}^{\text{sq}}$, respectively, noting that a change of reference frame must be made for each pair, in order for the expressions to apply. Closely separated pairs of squirmers and squirmers near the no-slip plate also experience a repulsive force $\bar{\bm{F}}^{\text{rep}}$ parallel to the vector joining their centres (or perpendicular to the wall). This follows the same functional form as in Eq.~\eqref{repulsive}, but in principle we allow squirmer-squirmer and squirmer-wall interactions to have different strengths $F_0$ and interaction ranges $\tau^{-1}$.

There are several forces and torques which do not require pairwise geometries. Every squirmer is subject to a propulsive force parallel to its orientation vector, $\bm{p}_i$, according to
\begin{align}
\bar{\bm{F}}_i^{\text{prop}} = 6 V_s \bm{p}_i,
\end{align}
where $V_s = 2B_1/3$ is the swimming speed of a solitary squirmer. Each squirmer also experiences a gravitational torque in the $z$-direction due to bottom-heaviness:
\begin{align}
\bar{\bm{T}}_i^{\text{grav}} = -\frac{1}{\pi} V_s G_{bh} \bm{p}_i \times \bm{g},
\end{align}
where $\bm{g} = -g (\sin \alpha, \cos \alpha, 0)$ and $g$ is the strength of gravity. Finally, the effect of the background shear flow is to exert a hydrodynamic force and torque on each squirmer as follows:
\begin{align}
\bar{\bm{F}}_i^{\infty} &= 6 \gamma y_i \bm{e}_x, \\
\bar{\bm{T}}_i^{\infty} &= -4 \gamma \bm{e}_z,
\end{align}
where $y_i$ is the $y$ coordinate of the $i^{\text{th}}$ squirmer. Before proceeding, it is instructive to consider the dynamics of the system if interactions between squirmers and the bounding plates are completely neglected. Under these conditions, the matrix system Eq.~\eqref{lubrication_resistance2} reduces to the following:
\begin{align}
\bm{R}^{\text{drag}} \cdot
\left( \begin{array}{c}
\bm{V} \\
a \bm{\omega} \\
\end{array} \right) &= 
- \left( \begin{array}{c}
\bar{\bm{F}}^{\text{prop}} + \bar{\bm{F}}^{\infty} \\
\bar{\bm{T}}^{\text{grav}}  + \bar{\bm{T}}^{\infty} \\
\end{array} \right). \label{lubrication_resistance3}
\end{align}
Since $\bm{R}^{\text{drag}}$ is diagonal, it is easily inverted, and the following results are obtained:
\begin{subequations} \label{solitary}
\begin{eqnarray}
& \bm{V}_i = V_s \bm{p}_i + \gamma y_i \bm{e}_x \label{decoupled_V}, \\
& a \bm{\omega} = -\frac{1}{8 \pi} V_s G_{bh} (\bm{p}_i \times \bm{g})  - \frac{1}{2} \gamma \bm{e}_z. \label{decoupled_omega} 
\end{eqnarray}
\end{subequations}
Equation~\eqref{decoupled_V} reveals that a solitary squirmer will swim at speed $V_s$ in the direction of its orientation, and be advected by the background shear flow in the $x$-direction. Similarly, the orientation of the squirmer will evolve according to the gravitational torque and background vorticity \citep{Jeffery1922}. In addition to these solitary dynamics (Eq.~\eqref{solitary}), the complete resistance formulation in Eq.~\eqref{lubrication_resistance2} includes hydrodynamic and steric effects through pairwise interactions, but only for sufficiently close squirmers.

The ability for ``solitary'' squirmers -- i.e. squirmers with no neighbours within lubrication range -- to propel themselves is a critical advancement of the present model. This maintains realistic behaviour at low areal fractions of the suspension, when it is quite feasible that a squirmer $i$ will have $\epsilon_{ij}>1 \ \forall \ j$, and ensures that the matrix system in Eq.~\eqref{lubrication_resistance2} is well conditioned for all configurations. 

The dynamics of the sheared squirmer suspension are calculated numerically by solving Eq.~\eqref{lubrication_resistance2} with the \textsc{Matlab} solver {\it ode15s}. The squirmers, initially distributed on a hexagonal close-packed array with mean spacing $\epsilon_0$, are each subject to a random translational perturbation within a disk of radius $\epsilon_0/2$, ensuring that cells are non-overlapping. Initially, the squirmers' orientations are taken to be random.

\subsection{Calculation of shear viscosity}

For the configuration presented in Fig.~\ref{fig10}, determining the effective suspension viscosity requires calculating the wall shear stress on each of the no-slip plates at $y = \pm H/2$. We emphasise that only squirmers which are sufficiently close to the walls to facilitate lubrication interactions will contribute to the wall shear stress. Of the full set of squirmers $S = \{ 1, 2, \ldots, N \}$, the following subsets can be identified:
\begin{align}
S_{\text{T}} &=\{ i \in \mathbb{N}  \, \big| \, | \tfrac{H}{2} - (y_i+a) |/a < 1 \}, \\
S_{\text{B}} &=\{ i \in \mathbb{N}  \, \big| \, | (y_i-a)+\tfrac{H}{2} |/a < 1 \}.
\end{align}
The sets $S_{\text{T}}$ and $S_{\text{B}}$ identify squirmers that have a clearance of less than $a$ with the top and bottom plates respectively (i.e. $\epsilon <1$), and therefore whose behaviour contributes to the lubrication forces and torques. The $x$-component of the force on the bottom plate is given by 
\begin{align}
\bar{F_x}^{\text{B}}  = F_x^{\text{B}}/ (\eta_0 \pi a) &= \sum_{i \in S_{\text{B}}} \big( F_x^{\text{sq}(i)} + F_x^{\text{trans}(i)} + F_x^{\text{rot}(i)} \big), \label{F_x_B}
\end{align}
where the three terms
\begin{subequations} \label{F_x_B_terms}
\begin{align}
F_x^{\text{sq}(i)} &= -\frac{6}{5} a \gamma Sq \bigg[1 - \beta (\bm{p}_i \cdot \bm{e}_y)  \bigg] (\bm{p}_i \cdot \bm{e}_x) \log \epsilon_i^{\text{B}}, \label{F_x_sq} \\
F_x^{\text{trans}(i)} &= -\frac{16}{5} \bigg[ \frac{H}{2} \gamma + \bm{V}_i \cdot \bm{e}_x \bigg] \log \epsilon_i^{\text{B}},  \label{F_x_trans}  \\ 
F_x^{\text{rot}(i)} &= -\frac{4}{5}  a \omega_i \log \epsilon_i^{\text{B}},  \label{F_x_rot} 
\end{align}
\end{subequations}
represent the $x$-component of the force on the plate, due to the squirming motion of the $i^{\text{th}}$ sphere, the difference between the squirmer velocity, $\bm{V}_i$, and the plate velocity, $-(H \gamma/2) \bm{e}_x$, and the angular velocity of the squirmer, respectively. In Eq.~\eqref{F_x_sq} we have made use of the fact that $B_1 = \frac{3}{2} a \gamma Sq $ and $B_2 = \beta B_1$. Similar expressions exist to calculate the tangential force, $\bar{F_x}^{\text{T}}$, on the top plate. Here, the value $\epsilon_i^{\text{B}}a$ represents the minimum clearance between the $i^{\text{th}}$ squirmer and the bottom plate. The repulsive force acts normal to the wall, and therefore does not contribute to the shear force. Although the background shear and gravitational torques influence the motion of the squirmers, their combined effects are encapsulated in Eqs.~\eqref{F_x_trans} and \eqref{F_x_rot}.

In dimensional form, the tangential shear stress on the top and bottom plates are given by $\sigma^{\text{T}} = \eta_0 \pi a \bar{F_x}^{\text{T}} / (L\delta)$ and $\sigma^{\text{B}} = \eta_0 \pi a \bar{F_x}^{\text{B}} / (L\delta)$, respectively, where $L = (2+\epsilon_0)a n_x$ and $\delta = 2.1a$ is the thickness of the monolayer (see Section~\ref{sec:2}). The effective bulk viscosity is therefore equal to
\begin{align}
\eta &=  \frac{\sigma^{\text{B}} - \sigma^{\text{T}}}{2\gamma} = \frac{\eta_0 \pi a}{2\gamma L\delta} \big( \bar{F_x}^{\text{B}} - \bar{F_x}^{\text{T}} \big).
\end{align}
The above expression utilises the mean value of the shear stress across both plates. The excess apparent viscosity is therefore equal to 
\begin{align}
\frac{\eta-\eta_0}{\eta_0} &= \frac{\pi \big( \bar{F_x}^{\text{B}} - \bar{F_x}^{\text{T}} \big)}{4.2 a \gamma (2+\epsilon_0)n_x} - 1. \label{lubrication_viscosity}
\end{align}
Upon first glance, it appears that the viscosity depends on the system size through the denominator in Eq.~\eqref{lubrication_viscosity}. Although the number of squirmers interacting with the wall would depend on suspension micro-structure, to leading order the number of terms in Eq.~\eqref{F_x_B} is proportional to $n_x$. Moreover, the terms in Eq.~\eqref{F_x_B} have the dimensions of velocity, matching the dimensions of $a \gamma$ in Eq.~\eqref{lubrication_viscosity}. In what follows, we will compare the results of Section~\ref{sec:2} using Eq.~\eqref{excessvis}, with the present lubrication formulation Eq.~\eqref{lubrication_viscosity}.

We should note here that the lubrication-theory-based (LT) method does not give an obvious way to compute normal stresses, so the results that follow will concentrate on predicting the shear viscosity.

\subsection{Results for non-bottom-heavy squirmers}

\begin{figure}
\begin{center}
\centerline{\includegraphics[width=0.96\textwidth]{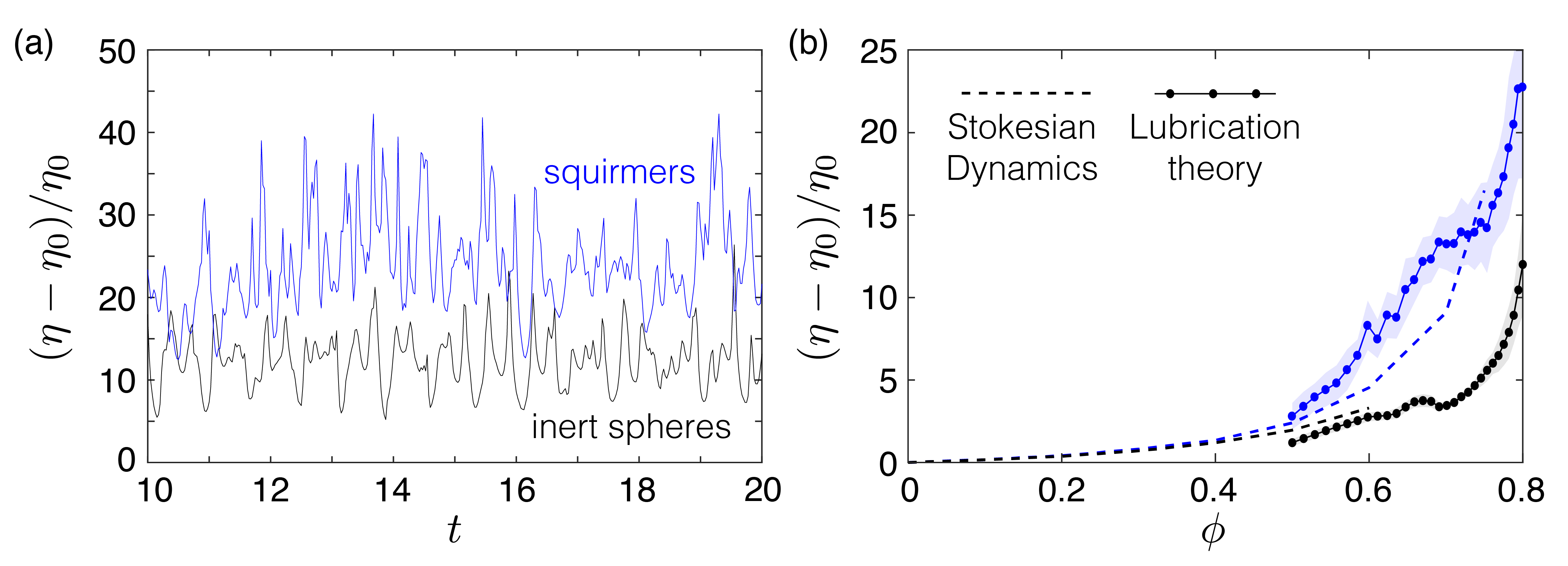}}
\caption{Excess apparent viscosity of a suspension of 132 spheres. (a) Time-dependent viscosity for inert spheres (black) and squirmers (blue) with $Sq = 1$ and $\beta = 1$. In both cases, the areal fraction is $\phi=0.80$. (b) Time-averaged suspension viscosity as a function of areal fraction $\phi$. Results are shown for lubrication simulations (cirlces), alongside the Stresslet method of Section~\ref{sec:2} (dashed). Shading represents the standard deviation of the time-dependent viscosity.}
\label{fig11}
\end{center}
\end{figure}

We first calculate the excess apparent viscosity in the absence of gravity. Suspensions of $N=132$ spheres with an areal fraction of $\phi=0.80$ were simulated over $t \in [0, 150]$, for both active squirmers ($Sq=1$, $\beta=1$) and passive spheres ($Sq=0$). The suspension viscosity can be calculated at every time-step using Eq.~\eqref{lubrication_viscosity}, the results of which are plotted in Fig.~\ref{fig11}(a). It is evident that the squirmer suspension (blue) has a higher mean viscosity than the suspension of passive spheres (black), and also exhibits greater excursions from the mean value.

The contribution to the mean squirmer suspension viscosity (blue curve in Fig.~\ref{fig11}(b)) arising purely from linear and angular velocities is $\eta/\eta_0-1 = 20.6$, approximately 90\% of the total mean viscosity ($\eta/\eta_0-1 = 22.8$). This is still considerably higher than the mean value for the passive sphere case $\eta/\eta_0-1 = 12.0$. Despite the propensity for active swimmers to redistribute themselves, the interior interactions still give rise to a higher suspension viscosity.

The areal fraction of cells was systematically varied in the range $0.5 < \phi < 0.8$ for both squirmers and passive spheres. A sufficiently long averaging window ($10<t<150$) was taken when calculating the time-averaged suspension viscosity. Figure~\ref{fig11}(b) shows the results of the lubrication simulations (circles) together with the findings using Stokesian dynamics (dashed, cf. Fig.~\ref{fig2}). There is good quantitative agreement between the complementary simulation methods across all values of $\phi$ studied.

\begin{figure}
\begin{center}
\centerline{\includegraphics[width=0.96\textwidth]{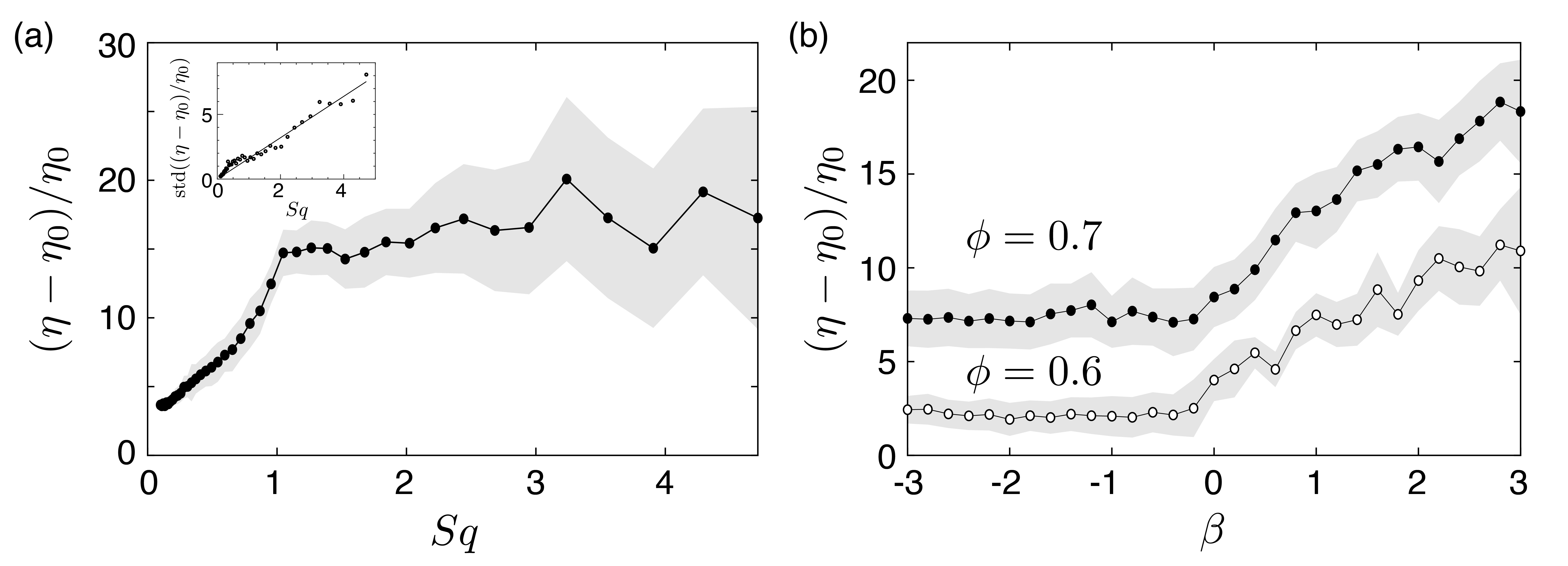}}
\caption{Effect of swimming properties on the mean suspension viscosity. (a) Excess apparent viscosity as a function of normalised swimming speed, $Sq$ (with $\beta=1$, $\phi=0.7$). Inset shows that the standard deviation in the viscosity increases with $Sq$. (b) Excess apparent viscosity as a function of swimming mode $\beta$ (with $Sq=1$) for $\phi=0.6$ (open) and $0.7$ (closed). Shading represents the standard deviation of the time-dependent viscosity.}
\label{fig12}
\end{center}
\end{figure}

The effect of the squirmers' swimming properties was also studied. The two dimensionless parameters are $Sq = V_s/{a \gamma}$, the swimming speed relative to background shear rate, and $\beta=B_2/B_1$, which effectively controls the stresslet sign and strength. Figure~\ref{fig12}(a) for $\beta = 1$ shows that the suspension viscosity increases for small $Sq$ before plateauing around $Sq=1$. Increasing $Sq$ further does not result in a noticeable shift in the viscosity, on average.

The dependence of the suspension viscosity on the swimming mode $\beta$ is shown in Fig.~\ref{fig12}(b) for two different areal fractions. The viscosity is higher for $\phi=0.7$ than for $\phi=0.6$ for all values of $\beta$ studied. The viscosity for pushers ($\beta < 0$) does not vary significantly with $\beta$, whereas for pullers ($\beta > 0$) the viscosity increases dramatically with $\beta$. This increase is much more marked than the Stokesian dynamics findings (see Fig.~\ref{fig4}(a)). Cross channel probability distributions for the squirmer positions reveal that pullers spend more time near the boundaries than pushers do, and this likely has a corresponding effect on the suspension viscosity.

\subsection{Results for bottom-heavy squirmers}

The effect of bottom-heaviness is to provide an external torque on each squirmer, which tends to reorient the cell in the gravitational field (see Fig.~\ref{fig10}). In the absence of hydrodynamic interactions or external shear, the orientation of each squirmer $\bm{p}_i$ will become anti-aligned with the gravity vector $\bm{g}$ over a timescale that is inversely proportional to the normalised strength of gravity, $G_{bh}$. A series of simulations were performed in which the strength of gravity and the angle with respect to the shear flow, were independently varied.

\begin{figure}
\begin{center}
\centerline{\includegraphics[width=\textwidth]{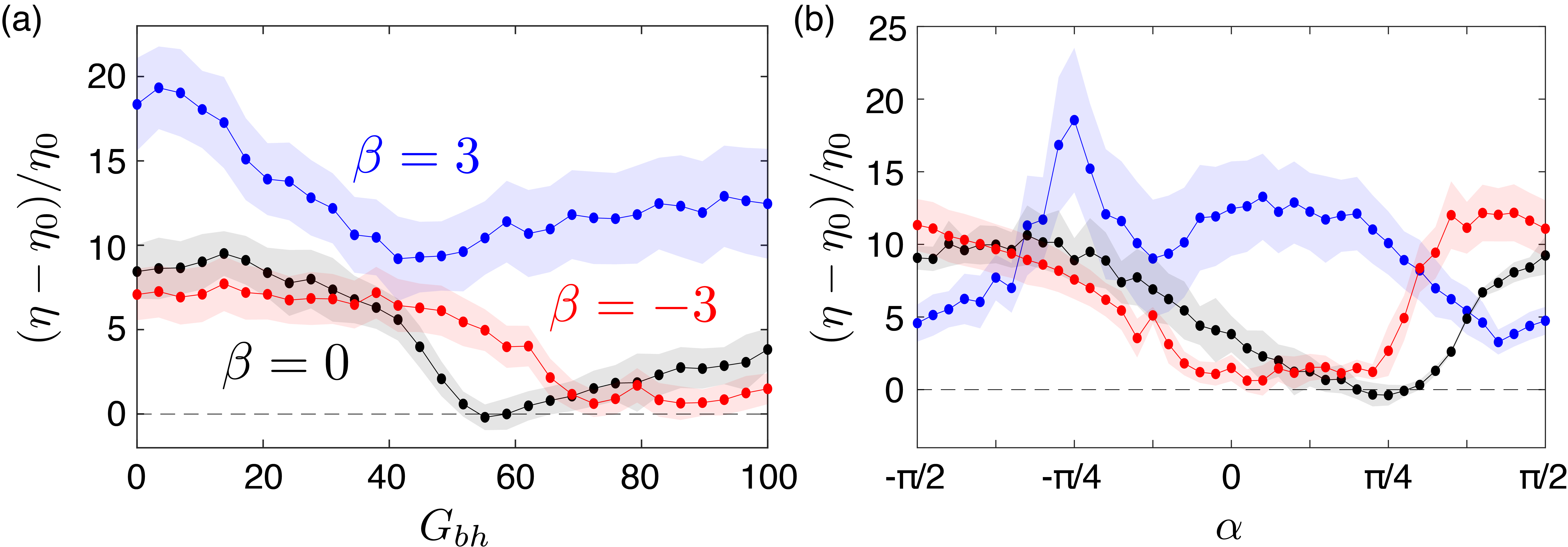}}
\caption{Effect of bottom-heaviness on the excess apparent viscosity ($Sq=1$, $\phi=0.7$). (a) The excess apparent viscosity as a function of $G_{bh}$ (with $\alpha=0$). (b) The effect of changing the gravity angle $\alpha$ (with $G_{bh}=100$). Results for pullers ($\beta=3$), pushers ($\beta=-3$), and neutral squirmers ($\beta=0$) are shown. Shading represents the standard deviation of the time-dependent viscosity.}
\label{fig13}
\end{center}
\end{figure}

Figure~\ref{fig13}(a) illustrates the excess apparent viscosity as a function of $G_{bh}$ for three different squirming modes, $\beta$ (all with $Sq=1$, $\phi=0.7$, $\alpha=0$). For small $G_{bh}$, the ordering matches the results of Fig.~\ref{fig12}(b), since hydrodynamic effects alone result in the strongest accumulation of pullers near the wall. As $G_{bh}$ is increased, the viscosity differences become more pronounced, with both neutral squirmers and pushers exhibiting $\eta/\eta_0-1 \approx 0$, i.e. no enhancement over the solvent viscosity. The internal squirmer motions are similar to those computed by the SD method: pullers with high $G_{bh}$ generate near-jamming aggregates. As the gravitational angle, $\alpha$, is varied (for fixed $G_{bh}=100$), the curves exhibit cross-over points (see Fig.~\ref{fig13}(b)). For $-\pi/4 \lesssim \alpha \lesssim \pi/4$, puller suspensions exhibit the greatest viscosity. However, for $\alpha \gtrsim \pi/4$ and $\alpha \lesssim -\pi/4$, pusher suspensions have a greater viscosity. In Fig.~\ref{fig13}(b), the large peak in the $\beta=3$ curve around $\alpha=-\pi/4$, corresponds to a situation where the (clockwise) viscous torques due to background shear balance the (counterclockwise) gravitational torques for upward-swimming squirmers. This results in a system which is, to leading order, equivalent to the $G_{bh}=0$ case (1\% difference between respective viscosities).

The dependence on $\alpha$ can be understood in terms of the fluid mechanics of propulsion in each case. For $\alpha \approx 0$, squirmers will tend to align vertically in the shear flow (i.e. perpendicular to the channel), so that their poles are closest to the walls. Moreover, since squirmers swim upwards, the lubrication interactions will be dominated by anterior poles interacting with the upper plate. Under these conditions, pullers will be drawn closer to the wall than pushers, and so will impart a greater lubrication drag on the top plate. For $\alpha \approx \pi/2$, gravity will tend to align squirmers parallel to the walls, so that equatorial regions of the squirmers are in the lubrication zones. The movement of fluid away from the poles in the case of pushers, means that under these circumstances, they will be drawn closer to the translating plates than their puller counterparts, and therefore exhibit a higher viscosity.

\section{Discussion} \label{sec:4}

The main focus of this paper has been to calculate the particle stress tensor in a concentrated suspension of spherical squirmers (modelling swimming cells) in a planar monolayer exposed to a uniform shear flow, over a wide range of parameter values, in the hope that the union of such results could act in place of an analytical constitutive relation, which appears unlikely to be achievable. For non-bottom-heavy squirmers, the particle stress tensor is symmetric and the rheology can be represented in terms of a single effective shear viscosity $\eta$, together with relatively small normal stress differences (Fig.~\ref{fig3}). The suspension is non-Newtonian because of these and the fact that $\eta$ depends on the value of the shear rate $\gamma$. However, for bottom-heavy squirmers the particle stress tensor is asymmetric ($\eta_{xy} \neq \eta_{yx}$) and exhibits significant normal stress differences: the suspension is clearly non-Newtonian. We have principally investigated how the time average of $\eta$ ($= (1/2)(\eta_{xy} + \eta_{yx})$ when the two off-diagonal components are different) varies with the areal fraction $\phi$, the squirming parameter $\beta$, the ratio $Sq$ of swimming speed to a typical speed of the shear flow, the bottom-heaviness parameter $G_{bh}$ (Eq.~\eqref{Gbh}), the angle $\alpha$ that the shear flow makes with the horizontal, and the two parameters $F_0$ and $\tau$ that define the repulsive force that is required computationally to prevent the squirmers from overlapping when their distance apart is too small.

We have employed two different numerical methods, Stokesian dynamics (SD), which takes account of all cell-cell interactions, including those between distant cells although high-order multipoles are neglected, and a lubrication-theory-based method (LT) that ignores all cell-cell interactions except those between a cell and its closest neighbours, which are calculated using the approximations of lubrication theory. Both computations are started from given initial conditions and results are taken when the effective viscosity has reached a statistically steady state. Both also use periodic boundary conditions in $x$. In the SD method, the underlying shear flow is imposed by applying a $y$-dependent translation to all the spheres, so that those at $y = L$, say, are displaced in the $x$-direction by an amount $L\gamma t$ relative to those at $y=0$; periodic boundary conditions in $y$ can then be applied. The particle stress tensor is given by the average over all spheres of the stresslet of an individual sphere (Eq.~\eqref{Stress_tensor_eqn}). In the LT method, the shear is applied by translating two infinite planes at $y = \pm H/2$ parallel to each other with velocity $\pm H\gamma/2$, and the suspension is set into motion by the viscous shear stresses $ \sigma^{\text{B}}$ (bottom wall) and $\sigma^{\text{T}}$ (top wall) exerted on them by the spheres nearest to those planes. The effective viscosity is $\eta = (\sigma^{\text{B}} - \sigma^{\text{T}})/2\gamma$. Thus the method of computing $\eta$ is very different between the two methods. They would give the same values only if most of the relevant physics took place in the interior of the suspension and not at the $y$-boundaries. In other words, although the only forces to be calculated in the LT method are the shear stresses at the walls, the viscosity depends on the internal interactions which drive changes to the configuration of the whole suspension by which the behaviour of the cells near the boundary are determined.

As shown in Fig.~\ref{fig11}(b), there is good agreement between the values of $\eta$ derived by the two methods for non-bottom-heavy squirmers, over the areal fraction range $0.5 < \phi < 0.75$.  Since the SD method takes into account multi-body interactions on top of the pairwise lubrication interactions, the agreement indicates that the relevant interactions are basically pairwise in the regime $0.5 < \phi < 0.75$ and confirms the basic validity of the lubrication theory approach. The variation of $\eta$ with various parameters also shows qualitative agreement between the two methods, though not good quantitative agreement: compare Figs.~\ref{fig5}(a) and \ref{fig12}(a) for $\eta(Sq)$, Figs.~\ref{fig6}(a) and \ref{fig13}(a) for $\eta(G_{bh})$, and Figs.~\ref{fig6}(b) and \ref{fig13}(b) for $\eta(\alpha)$. In particular, the variation of $\eta$ with $\beta$ for pullers (positive $\beta$) is much more marked according to the LT method (Fig.~\ref{fig12}(b)), even for non-bottom-heavy squirmers. To understand this, consider the respective interactions of pushers ($\beta<0$) and pullers ($\beta>0$) with a translating wall. The cells oriented towards the wall will have a tendency to reside there longer than cells pointing away from the wall, though this interpretation can obviously break down in various situations. It follows that the anterior hemispheres of the squirmers are most likely to constitute the lubrication interaction with the wall. As squirmers are ``tilted'' clockwise by the translating boundary, pullers will more strongly oppose the motion of the plate, thereby leading to a higher viscosity. This mechanism partly overrides the similarity of the internal configurations; it is a new physical mechanism, which, unlike those considered in previous studies, is not based on cell elongation.

It was seen in Fig.~\ref{fig2}(b) that the contribution to the effective viscosity of the repulsive forces between squirmers becomes larger at high volume fraction ($\phi \ge 0.6$) than the contribution from hydrodynamic forces. The results in that figure were obtained for $\beta = 1$, $Sq = 1$, using our `standard' values of repulsive force parameters, $F_0 = 10, \tau = 100$ ($F_0$ controls the strength of the repulsive force, whereas $1/\tau$ represents the distance over which the force plays a role). The dependence of the excess effective viscosity on the repulsive forces, calculated using the SD and LT methods, is shown in Table~\ref{table_F_repulsive}. The viscosity is only slightly affected by the parameters $F_0$ and $\tau$. On the other hand, the difference between the squirmers and inert spheres is pronounced, and of approximately the same magnitude, across all force combinations studied, indicating that the results are not critically dependent on the values of these parameters. The difference between squirmers and inert spheres is presumably induced by two mechanisms: a) the surface squirming velocity generates a direct contribution to the stresslet, and b) the squirming motion determines the suspension microstructure in the first place.

The magnitude of the repulsive force has a similar moderate influence on the apparent viscosity in the LT simulations. Both the magnitude $F_0$, and characteristic length scale $1/\tau$, of the repulsive force influence the spacing, $\epsilon$, between the bounding walls and the squirmers adjacent to them. Since the effective viscosity is determined solely by lubrication interactions which scale as $\log \epsilon$ (see Eq.~\eqref{F_x_B}), the repulsive force does influence the measured viscosity slightly. However, as in the SD method, we emphasise that the differences between the passive spheres and the squirmers prevail regardless of the specific choices of $F_0$ and $\tau$. 

\begin{table}
    \centering
    \vspace{-10pt}
    \caption{Effect of the repulsive force on the excess apparent viscosity $(\eta -\eta_0)/\eta_0$. Particles are non-bottom-heavy and the areal fraction is $\phi = 0.5$. Results are shown for inert spheres and squirmers (with $\beta=1$, $Sq=1$), for both SD and LT methods.} \label{table_F_repulsive}
    \begin{tabular}{c c c c c }
    \hline
     & \multicolumn{2}{c}{Stokesian dynamics (SD)} & \multicolumn{2}{c}{Lubrication theory (LT)}
     \vspace{4pt} \\
    $(F_0, \tau)$ & {\it ~~inert spheres~~} & {\it ~~squirmers~~} & {\it ~~inert spheres~~} & {\it ~~squirmers~~} \\
    \hline
    $(10/3, 100)$ & 1.98 & 2.41 & 1.32 & 2.86 \\
    $(10, 100)$ & {\bf 1.96} & {\bf 2.40} & {\bf 1.26} & {\bf 2.87} \\
    $(30, 100)$ & 1.87 & 2.28 & 1.18 & 2.92 \\
    $(10, 100/3)$ & 1.76 & 2.25 & 0.98 & 3.47 \\
    $(10, 300)$ & 2.01 & 2.68 & 1.36 & 2.53 \\
    \end{tabular}
\end{table}

Although the present paper is mainly concerned with the mean rheological properties, we also analysed time-dependent viscosity of the suspensions (see for example Fig.~\ref{fig11}(a)). The squirmer suspensions always exhibit a higher {\it mean} viscosity than the passive sphere case. Furthermore, the fluctuations about the mean, shown explicitly in Fig.~\ref{fig11}(a) and displayed as confidence intervals in subsequent figures, increase with areal fraction $\phi$. It is appropriate to note here that the maximum packing fraction for spheres in a monolayer is $\phi \approx 0.91$.

The model active suspension analysed in this paper is extremely idealised: a planar array, of identical, non-colloidal, spherical cells, which swim through the fluid by squirming with an unchanging distribution of tangential velocity on their surfaces. Their concentration (areal fraction $\phi$) is high (up to 0.8). It follows that there is not much previous literature against which the results can be compared, either computational or, especially, experimental. As discussed in Section 2.2, the only monolayer predictions of suspension rheology that we could find are those of \cite{SinghNott:2000}, who considered passive spheres and predicted a greater increase with $\phi$ of effective viscosity than found in this paper, as shown in Fig.~\ref{fig2}(a), and our own earlier work on squirmers at lower values of $\phi$ \citep{Ishikawa:2007rheology}. There have been predictions of suspension rheology in three-dimensional flows, using concentrated suspensions of rigid spheres \citep{Sierou:2002}, and experiments on dilute suspensions of swimming bacteria \citep{Lopez:2015}, but even in three dimensions there appear to be no findings on concentrated suspensions of micro-swimmers, apart from \cite{Ishikawa:2007rheology}. The computational work referred to here was conducted using Stokesian dynamics; analysing the same system using lubrication theory alone has been tried only by \cite{Leshansky:2005}, for inert spheres (in three dimensions) -- and this was referred to only in a footnote -- and \cite{BrumleyPedley:2019}, for squirmers, but not in order to predict the rheology.

The absence of previous relevant work, however, means that there is plenty of scope for future research. We would like to extend the LT method to three dimensions, where it might be quicker and easier to use than SD, enabling predictions of 3D suspension behaviour for a range of parameter values, building up a graphical representation of a suspension's constitutive relation, as we have tried to do here in 2D. A more complete description of the rheology, however, either in 2D or  3D, will require a quantitative representation of pressure gradients, which are absent in our current simulations since the particle stress tensor, derived from Eq.~\eqref{Stress_tensor_eqn}, is traceless and does not contribute to the pressure $P$ in Eq.~\eqref{bulk_stress_tensor}, though it does determine the normal stress differences. The presence of walls, for a suspension in a channel or pipe, introduces a particle pressure distinct from the average bulk pressure, because of the forces exerted by the particles that interact hydrodynamically with the walls. A clear explanation of this effect is given by \cite{Guazzelli:2018}. \cite{SinghNott:2000} computed normal stresses and pressure in their SD simulations of a suspension of inert spheres in a channel, and we would expect to follow their lead for squirmers.   

Other possible extensions to this work include consideration of particles of different shapes, such as prolate spheroids, to represent bacteria \citep{Saintillan:2018}, or squirmers that rotate about their axes, to represent \emph{Volvox} \citep{Pedley2016Volvox}. Either of these extensions would affect the dynamics of solitary squirmers, and hence their trajectories in shear flows and suspensions. At higher areal fractions reminiscent of granular media, shape would influence the propensity of cells to pack together, and control their nematic order. Moreover, lubrication interactions with the bounding walls would also be influenced by organismal shape \citep{Manabe2020} or rotation. 

The current model assumes that all particles are identical and move deterministically. In any real suspension, however, stochastic factors play a part, because of (possibly small) differences between individuals' shapes and locomotory apparatuses. As long as the differences are small, perturbation theory may be feasible, but since the motions of the particles in a  suspension become effectively random after a small number of collisions (see \cite{Ishikawa:2007diffusion}) this is unlikely to be profitable.  

Finally, although $G_{bh}$ is designed to mimic gravitactic micro-organisms such as {\it Volvox}, the analysis could also be applied to other systems in which an external field tends to align the cells (e.g. phototaxis, magnetotaxis).\\

\begin{flushleft}
{\large {\bf Acknowledgments}}
\end{flushleft}

T.I. was supported by the Japan Society for the Promotion of Science Grant-in-Aid for Scientific Research (JSPS KAKENHI Grant No. 17H00853 and No. 17KK0080). T.I. performed computations in Advanced Fluid Information Research Center, Tohoku University. D.R.B. was supported by an Australian Research Council (ARC) Discovery Early Career Researcher Award DE180100911. D.R.B. performed simulations using The University of Melbourne's High Performance Computer {\it Spartan}. 

T.J.P. would like to pay tribute to the major influence that the late George Batchelor exerted on his own development and career in fluid mechanics; George was severally his PhD supervisor, his intellectual guide, his head of department, and his general mentor (for example selecting him as an Associate Editor of JFM), over a period of at least thirty years. He was an inspiration to us all.

\end{document}